\documentclass[12pt]{article}
\newif\iffigs\figstrue
\usepackage{latexsym}
\iffigs
  \input{epsf}
\else
\fi
\font\tenmsbm=msbm10 scaled 1200
\font\sevenmsbm=msbm9
\newfam\msbmfam
\textfont\msbmfam=\tenmsbm
\scriptfont\msbmfam=\sevenmsbm

\newtheorem{definizione}{Definition}[section]
\newcommand{\bdefi}{\begin{definizione}}
\newcommand{\edefi}{\end{definizione}}
\makeatletter
\@addtoreset{equation}{section}
\makeatother

\newcommand{\eqn}[1]{(\ref{#1})}
\newcommand{\ft}[2]{{\textstyle\frac{#1}{#2}}}
\newsavebox{\uuunit}
\sbox{\uuunit}
    {\setlength{\unitlength}{0.825em}
     \begin{picture}(0.6,0.7)
        \thinlines
        \put(0,0){\line(1,0){0.5}}
        \put(0.15,0){\line(0,1){0.7}}
        \put(0.35,0){\line(0,1){0.8}}
       \multiput(0.3,0.8)(-0.04,-0.02){10}{\rule{0.5pt}{0.5pt}}
     \end {picture}}
\newcommand {\unity}{\mathord{\!\usebox{\uuunit}}}

\def\veck{\stackrel{\rightarrow}{k}}

\def\IP{\relax{\rm I\kern-.18em P}}

\font\tenmsbm=msbm10 scaled 1200
\font\sevenmsbm=msbm9
\newfam\msbmfam
\textfont\msbmfam=\tenmsbm
\scriptfont\msbmfam=\sevenmsbm

\def\dslash{\hbox{\ooalign{$\displaystyle\partial$\cr$/$}}}
\def\Aslash{\hbox{\ooalign{$\displaystyle A$\cr$\hspace{.03in}/$}}}
\def\Aplusslash{\hbox{\ooalign{$\displaystyle A^+$\cr$\hspace{.03in}/$}}}
\def\Aminusslash{\hbox{\ooalign{$\displaystyle A^-$\cr$\hspace{.03in}/$}}}

\def\Zminusslash{\hbox{\ooalign{$\displaystyle Z^-$\cr$\hspace{.03in}/$}}}

\def\nslash{\nabla\!\!\!\!/}
\def\inbar{\vrule height1.5ex width.4pt depth0pt}
\def\IC{\relax\,\hbox{$\inbar\kern-.3em{\rm C}$}}
\def\IG{\relax\,\hbox{$\inbar\kern-.3em{\rm G}$}}
\def\IB{\relax{\rm I\kern-.18em B}}
\def\ID{\relax{\rm I\kern-.18em D}}
\def\IL{\relax{\rm I\kern-.18em L}}
\def\IF{\relax{\rm I\kern-.18em F}}
\def\IH{\relax{\rm I\kern-.18em H}}
\def\II{\relax{\rm I\kern-.17em I}}
\def\IN{\relax{\rm I\kern-.18em N}}
\def\IP{\relax{\rm I\kern-.18em P}}
\def\IQ{\relax\,\hbox{$\inbar\kern-.3em{\rm Q}$}}
\def\bfzero{\relax\,\hbox{$\inbar\kern-.3em{\rm 0}$}}
\def\IK{\relax{\rm I\kern-.18em K}}
\def\IG{\relax\,\hbox{$\inbar\kern-.3em{\rm G}$}}
 \font\cmss=cmss10 \font\cmsss=cmss10 at 7pt
\def\IR{\relax{\rm I\kern-.18em R}}
\def\IGam{\relax{{\rm I}\kern-.18em \Gamma}}
\def\ZZ{\relax\ifmmode\mathchoice
{\hbox{\cmss Z\kern-.4em Z}}{\hbox{\cmss Z\kern-.4em Z}}
{\lower.9pt\hbox{\cmsss Z\kern-.4em Z}}
{\lower1.2pt\hbox{\cmsss Z\kern-.4em Z}}\else{\cmss Z\kern-.4em
Z}\fi}
\def\bfone{\relax{\rm 1\kern-.35em 1}}

\def\LL{{\cal L}}
\def\a{\alpha}
\def\b{\beta}
\def\d{\delta}
\def\l{\lambda}

\def\g{\gamma}

\def\s{\sigma}
\def\e{\epsilon}
\def\o{\omega}
\font\cmss=cmss10 \font\cmsss=cmss10 at 7pt

\def\inbar{\vrule height1.5ex width.4pt depth0pt}
\def\IC{\relax\,\hbox{$\inbar\kern-.3em{\rm C}$}}
\def\bfzero{\relax\,\hbox{$\inbar\kern-.3em{\rm 0}$}}
\def\bfone{\relax{\rm 1\kern-.35em 1}}

\def\tilde{\widetilde}

\def\IE{\relax{{\rm I\kern-.18em E}}}

\def\IGam{\relax{{\rm I}\kern-.18em \Gamma}}

\def\bet{\begin{tabular}}
\def\eet{\end{tabular}}

\def\a{\alpha}
\def\b{\beta}
\def\l{\lambda}

\def\g{\gamma}

\def\s{\sigma}
\def\e{\epsilon}

\begin{document}
\begin{titlepage}
\begin{flushright}
hep-th/9905134\\
May 1999\\
\end{flushright}
\vskip 2cm
\begin{center}
{\Large \bf    $Osp(N|4)$ supermultiplets \\
\vskip  0.5 cm
as conformal superfields on   $\partial AdS_4$ and\\
\vskip  0.5 cm
 the generic form of $N=2,\, d=3$ gauge theories $^*{}^\dagger$ }\\
\vfill
{\large    Davide Fabbri,
 Pietro Fr\'e, Leonardo Gualtieri and Piet Termonia   } \\
\vfill
{
  Dipartimento di Fisica Teorica, Universit\'a di Torino, via P.
Giuria 1,
I-10125 Torino, \\
 Istituto Nazionale di Fisica Nucleare (INFN) - Sezione di Torino,
Italy \\
}
\end{center}
\vfill
\begin{abstract}
In this paper we fill a necessary gap in order to realize the explicit
comparison between the Kaluza Klein spectra of supergravity
compactified on $AdS_4 \times X^7$ and superconformal field theories
living on the world volume of M2--branes. On the algebraic side we
consider the superalgebra $Osp({\cal N}\vert 4)$ and we  study the
double intepretation of its unitary irreducible representations
either as supermultiplets of particle states in the bulk or as
conformal superfield on the boundary. On the lagrangian field theory
side we construct, using rheonomy rather than superfield techniques,
the generic form of an ${\cal N}=2,d=3$ gauge theory. Indeed the
superconformal multiplets are supposed to be composite operators in
a suitable gauge theory.
\end{abstract}
\vspace{2mm} \vfill \hrule width 3.cm
{\footnotesize
 $^*$ Supported in part by   EEC  under TMR contract
 ERBFMRX-CT96-0045}
\end{titlepage}
\section{Introduction}
\label{intro}
One of the most exciting developments in the recent history of string
theory has been the discovery of the holographic AdS/CFT
correspondence \cite{renatoine,maldapasto,townrenatoi,serfro1,serfro2,serfro3}:
\begin{equation}
\label{holography}
\mbox{SCFT on $\partial (AdS_{p+2})$} \quad \leftrightarrow \quad
\mbox{SUGRA on $ AdS_{p+2}$}
\end{equation}
between a   {\it $d=p+1$ quantum superconformal field theory} on
 the boundary of anti de Sitter space and {\it  classical Kaluza
Klein supergravity} \cite{Kkidea,freurub,round7a,squas7a,osp48},
\cite{kkwitten,noi321,spectfer,univer,freedmannicolai,multanna},
\cite{englert,biran,casher},
\cite{dafrepvn,dewit1,duffrev,castromwar,gunawar,gunay2}
emerging from compactification of either
superstrings  or M-theory  on the product space
\begin{equation}
AdS_{p+2} \, \times \, X^{D-p-2}\,,
\label{adsX}
\end{equation}
where $X^{D-p-2}$ is a $D-p-2$--dimensional compact Einstein manifold.
\par
The present paper deals with the case:
\begin{equation}
  p=2 \quad \leftrightarrow \quad d=3
\label{choice}
\end{equation}
and studies two issues:
\begin{enumerate}
  \item The relation between the description of unitary irreducible
representations
   of the  $Osp(2 \vert 4)$ superalgebra seen as off--shell conformal
  superfields in $d=3$ or as  on--shell particle supermultiplets in
  $d=4$ anti de Sitter space. Such double interpretation of the same
  abstract representations is the algebraic core of the AdS/CFT
  correspondence.
  \item The generic component form of an ${\cal N}=2$ gauge theory in three
  space-time dimensions containing the supermultiplet of an arbitrary
  gauge group, an arbitrary number of scalar multiplets in arbitrary
representations
  of the gauge group and with generic superpotential interactions.
  This is also an essential item in the discussion
of the AdS/CFT correspondence since the superconformal field theory
on the boundary is to be identified with a superconformal infrared
fixed point of a non abelian gauge theory of such a type.
\end{enumerate}
Before presenting our results we put them into perspective through
the following introductory remarks.
\subsection{The conceptual environment and our goals}
The logical path connecting the two partners in the above correspondence
(\ref{holography}) starts from considering a special instance of
classical $p$--brane solution of $D$--dimensional supergravity
characterized by the absence of a dilaton ($a=0$ in standard
$p$--brane notations) and by the following relation:
\begin{equation}
  \frac{d  {\tilde d} }{D-2} = 2
\label{specrel}
\end{equation}
between the dimension $d \equiv p+1$ of the $p$--brane world volume
and that of its magnetic dual $ {\tilde d} \equiv D-p-3$.
Such a solution is given by the following metric and  $p+2$ field strength:
\begin{eqnarray}
ds^2_{brane} & =& \left(1+\frac{k}{ r^{\tilde d} } \right)^{-  \frac { { \tilde
d}} {  (D-2)}}
\, dx^m \otimes dx^n \, \eta_{mn}
- \left(1+\frac{k}{ r^{\tilde d} } \right)^{ \frac {  {d}} {  (D-2)}}
\,  \left ( dr^2 + r^2 \, ds^2_{X}(y) \right )\,,\nonumber \\
F \equiv  dA &= &\lambda (-)^{p+1}\epsilon_{m_1\dots m_{p+1}} dx^{m_1}
\wedge \dots \wedge dx^{m_{p+1}}
\wedge \, dr \, \left(1+\frac{k}{r^{\tilde
d}}\right )^{-2} \, \frac{1}{r^{{\tilde d}+1}}\,.
\label{elem}
\end{eqnarray}
In eq. (\ref{elem}) $ds^2_{X}(y)$ denotes the Einstein metric on the
compact manifold $X^{D-p-2}$ and the $D$ coordinates have been
subdivided into the following subsets
\begin{itemize}
\item $x^m$ $(m=0,\dots ,p)$ are the coordinates on the
$p$--brane world--volume,
\item $r= \mbox{radial}$ $\oplus$ $y^x =\mbox{angular on $X^{D-p-2}$}$  $(x=D-
d+2,\dots ,D)$
are the coordinates  transverse to the brane.
\end{itemize}
In the limit $ r \to 0$ the classical brane metric $ds^2_{brane}$
approaches the following metric:
\begin{equation}
   \begin{array}{ccccc}
     ds^2 & = & \underbrace{\frac{r^{2 {\tilde d}/d}}{k^{2/d}} \,
     dx^m \otimes dx^n \,\eta_{m n }-\frac{k^{2/{\tilde d}}}{r^2} \, dr^2}
      & - &\underbrace{\phantom{\frac{1}{1^{1}}} k^{2/{\tilde d}} \, ds^2_X(y)
      \phantom{\frac{1}{1^{1}}}} \\
     \null & \null & AdS_{p+2} & \times  & X^{D-p-2}\,,\\
\end{array}
\label{rto0}
\end{equation}
that is easily identified as the standard metric on the product space
$AdS_{p+2} \times X^{D-p-2}$. Indeed it suffices to set:
\begin{equation}
  \rho=r^{{\tilde d}/d} \, k^{2(d+{\tilde d})/d^2}
\label{rearra}
\end{equation}
to obtain :
\begin{eqnarray}
ds^2_{brane} & \stackrel{r \to 0}{\approx} & k^{2/{\tilde d}}
    \,\left(ds^2_{AdS} -ds^2_X(y)\right)\,,
    \label{produ}\\
ds^2_{AdS} & =& \left( \rho^2 \, dx^m \otimes dx^n \,\eta_{m n }-
\frac{d\rho^2}{\rho^2} \right)\,,
\label{adsmet}
\end{eqnarray}
where (\ref{adsmet}) is the canonical form of the anti de Sitter
metric in solvable coordinates \cite{g/hpape}.
\par
On the other hand, for $r \to \infty$ the brane metric approaches the
limit:
\begin{equation}
\begin{array}{ccccc}
ds^2_{brane} &\stackrel{r\to \infty}{\approx} &
\underbrace{dx^m \otimes dx^n \,\eta_{m n
}} & - & \underbrace{dr^2 +r^2 \, ds^2_{X}(y)}\\
\null & \null & M_{p+1} & \null & C\left (X^{D-p-2}\right)\,,
\end{array}
\end{equation}
where $M_{p+1}$ denotes Minkowski space in $p+1$ dimensions while $C\left
(X^{D-p-2}\right)$ denotes the $D-p-1$ dimensional {\it metric cone} over the
{\it horizon
manifold} $X^{D-p-2}$. The key point is that (compactified) Minkowski space can
also
be identified with the boundary of anti de Sitter space:
\begin{equation}
  \partial \left(  AdS_{p+2} \right)  \equiv M_{p+1}\,,
\label{identifia}
\end{equation}
so that we can relate supergravity on $AdS_{p+2} \times X^{D-p-2}$ to
the {\it gauge theory} of a stack of $p$--branes placed in such a way
as to have the metric cone  as transverse space (see fig.\ref{cono1})
\begin{equation}
  C\left
(X^{D-p-2}\right) = \mbox{transverse directions to the branes.}
\label{transverse}
\end{equation}
\iffigs
\begin{figure}
\caption{The metric cone  $C\left (X^{D-p-2}\right)$ is transverse to the stack
of branes}
\begin{center}
\label{cono1}
\epsfxsize = 10cm
\epsffile{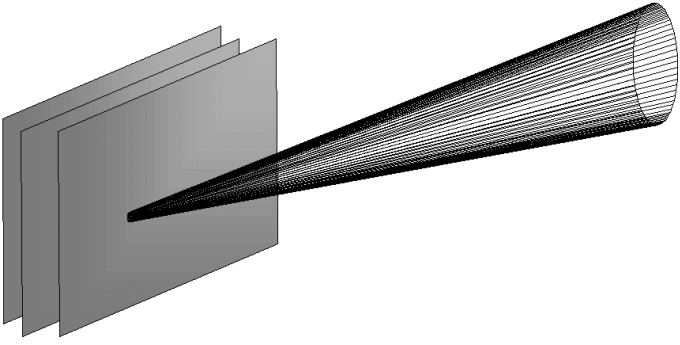}
\vskip  0.2cm
\hskip 2cm
\unitlength=1.1mm
\end{center}
\end{figure}
\fi
\par
\par
According to current lore on brane dynamics
\cite{branedyn1,branedyn2,branedyn3,branedyn4},
 if the metric cone $C(X^{D-p-2})$
can be reintepreted as some suitable resolution of an orbifold
singularity \cite{c/ga1,c/ga2,c/ga3,c/ga4}:
\begin{eqnarray}
  C(X^{D-p-2}) &=& \mbox{resolution of } \,
  \frac{R^{D-p-1}}{\Gamma}\,,\nonumber\\
  \Gamma & = & \mbox{discrete group},
\label{resolut}
\end{eqnarray}
then there are means to identify a {\it gauge theory} in $M_{p+1}$
Minkowski space with supersymmetry ${\cal N}$ determined by the
holonomy of the metric cone, whose structure and field content in the
ultraviolet limit
is determined by the orbifold $ {R^{D-p-1}}/{\Gamma}$. In the
infrared limit, corresponding to the resolution $C(X^{D-p-2})$, such a
gauge theory has a superconformal fixed point and defines the
superconformal field theory $SCFT_{p+1}$ dual to supergravity on
$AdS_{p+2} \times X^{D-p-2}$.
\par
In this general conceptual framework there are
three main interesting cases where the basic  relation
(\ref{specrel}) is satisfied
\begin{equation}
   \begin{array}{cccl}
     p=3 & D=10 &   AdS_5  \times  X^5 &\mbox{   $D3$--brane of type IIB
theory} \\
     p=2 & D=11 &   AdS_4  \times   X^7 &  \mbox{  $M2$--brane of M--theory} \\
     p=5 & D=11 &  AdS_7  \times   X^4  &\mbox{   $M5$--brane of M--theory} \
   \end{array}
\label{casi}
\end{equation}
The present paper focuses on the case of $M2$ branes and on the
general features of ${\cal N}=2$ superconformal field theories
in $d=3$. Indeed the final goal we are pursuing in a series
of papers is that of determining  the  three--dimensional superconformal field
theories
dual to compactifications of D=11 supergravity on $AdS_4 \times X^7$,
where the non spherical horizon $X^7$ is chosen to be one
of the four  homogeneous sasakian $7$--manifolds $G/H$:
\begin{equation}
  X^7 = \cases{
  \begin{array}{ccc}
    M^{1,1,1} & = & \frac{SU(3) \times SU(2) \times U(1)}{SU(2) \times U(1)
\times U(1)} \\
    Q^{1,1,1} &=& \frac{SU(2) \times SU(2) \times SU(2)}{ U(1) \times U(1) }
\\
    N^{0,1,0} & = & \frac{SU(3)}{U(1)}  \\
    V_{5,2} & = & \frac{SO(5)\times SO(2)}{SO(3) \times SO(2)}
  \end{array}}
\label{sasaki}
\end{equation}
that were classified  in the years 1982-1985 \cite{noi321,dafrepvn,castromwar}
when Kaluza Klein supergravity was very topical. The Sasakian
structure \cite{sask1,sask2,sask3,sask4}
of $G/H$ reflects its holonomy and is the property that
guarantees ${\cal N}=2$ supersymmetry  both in the bulk $AdS_4$ and on
the boundary $M_3$. Kaluza Klein spectra for $D=11$ supergravity
compactified on the manifolds (\ref{sasaki}) have already been
constructed \cite{m111spectrum} or are under construction \cite{loro52}
and, once the corresponding superconformal theory has been
identified, it
can provide a very important tool for comparison and discussion of
the AdS/CFT correspondence.
\subsection{The specific problems addressed and solved in this paper}
In the present paper we do not address the question of constructing
the algebraic conifolds defined by the metric cones $C(G/H)$
nor the identification of the corresponding orbifolds.  Here we do not
discuss the specific construction of the superconformal field theories
associated with the horizons (\ref{sasaki}) which
is postponed to future publications \cite{cesar}: we rather   consider
a more general problem that constitutes an intermediate and essential
step for the comparison between Kaluza Klein spectra and
superconformal field theories. As anticipated above, what we need is
a general translation vocabulary between the two descriptions
of  $Osp({\cal N} \vert 4)$ as the superisometry algebra in anti de
Sitter ${\cal N}$--extended $d=4$ superspace and as a superconformal
algebra in $d=3$. In order to make the comparison between
superconformal field theories and Kaluza Klein results explicit, such
a translation vocabulary is particularly essential at the level of
unitary irreducible representations (UIR).  On the Kaluza Klein side the
UIR.s  appear as supermultiplets of on--shell particle
states characterized by their square mass $m^2$  which, through well established
 formulae, is expressed as a quadratic form:
\begin{equation}
  m^{2} = c \, (E_0-a)(E_0 -b)
\label{massformu}
\end{equation}
in the {\it energy eigenvalue} $E_0$ of a compact $SO(2)$ generator,
by their spin $s$ with respect to the compact  $SO(3)$ {\it little group}
of their momentum vector and, finally, by a set of $SO({\cal N})$
labels. These particle states live in the bulk of $AdS_4$.
On the superconformal side the UIR.s appear instead as multiplets of primary
conformal operators constructed out of the fundamental fields of the
gauge theory. They are characterized by their conformal weight $D$,
their $SO(1,2)$ spin $J$ and by the labels of the $SO({\cal N})$
representation they belong to.
Actually it is very convenient to regard such multiplets
of conformal operators as appropriate conformally invariant superfields
in $d=3$ superspace.
\par
Given this, what one needs is a general framework to
convert data from one language to the other.
\par
Such a programme has been extensively developed in the case of the
$AdS_5/CFT_4$ correspondence between ${\cal N}=4$ Yang--Mills theory in
$D=4$, seen as a superconformal theory, and type IIB supergravity
compactified on $AdS_5 \times S^5$. In this case the superconformal
algebra is $SU(2,2\vert 4)$ and the relation between the two
descriptions of its UIR.s as boundary or bulk supermultiplets was
given, in an algebraic setup, by Gunaydin and collaborators
 \cite{gunaydinminiczagerman1,gunaydinminiczagerman2}, while the
corresponding superfield description was discussed in a series of
papers by Ferrara and collaborators
\cite{sersupm1,sersupm2,sersupm3,sersupm4}.
\par
A similar discussion for the case of the $Osp({\cal N} \vert 4)$
superalgebra was, up to our knowledge, missing so far. The present
paper is meant to fill the gap.
\par
There are relevant structural differences between the  superalgebra
${\bf G} = SU(2,2\vert {\cal N})$ and the superalgebra ${\bf G}= Osp({\cal N}
\vert 4)$
 but the basic strategy of papers
\cite{gunaydinminiczagerman1,gunaydinminiczagerman2} that
consists of performing a suitable {\it rotation} from a basis of
eigenstates of the maximal compact subgroup $SO(2) \times SO(p+1) \subset {\bf
G}$
to a basis of eigenstates of the maximal non compact subgroup
$SO(1,1) \times SO(1,p) \subset {\bf G}$
can be adapted. After such a rotation we  derive the $d=3$ superfield
description of the supermultiplets by means of a very simple and
powerful method based on the {\it supersolvable parametrization } of
anti de Sitter superspace \cite{torinos7}. By definition, anti de Sitter
superspace
is the following supercoset:
\begin{equation}
  AdS_{4\vert{\cal N} } \equiv \frac{Osp({\cal N} \vert 4)}{SO(1,3) \times
SO({\cal N})}
\label{ads4N}
\end{equation}
and has $4$ bosonic coordinates labeling the points in $AdS_4$ and $4
\times {\cal N}$ fermionic coordinates $\Theta^{\alpha i}$ that
transform as Majorana spinors under $SO(1,3)$ and as vectors under
$SO({\cal N})$. There are many possible coordinate choices for
parametrizing such a manifold, but as far as the bosonic submanifold is
concerned it was shown in \cite{g/hpape} that a particularly useful
parametrization is the solvable one where the $AdS_4$ coset is
regarded as a {\it non--compact solvable group manifold}:
\begin{equation}
  AdS_4 \equiv \frac{SO(2,3)}{SO(1,3)} = \exp \left [ Solv_{adS}
  \right]
\label{solvads}
\end{equation}
The solvable algebra $Solv_{adS}$  is spanned by the unique
non--compact Cartan generator $D$ belonging to the coset and
by three abelian operators $P_m$ ($m=0,1,2$) generating the
translation subalgebra in $d=1+2$ dimensions. The solvable
coordinates are
\begin{equation}
   \begin{array}{rclcrcl}
     \rho & \leftrightarrow  & D & ; & z^m & \leftrightarrow & P_m \
   \end{array}
\label{solvcord}
\end{equation}
and in such coordinates the $AdS_4$ metric takes the form
(\ref{adsmet}). Hence $\rho$ is interpreted as measuring the distance
from the brane--stack and $z^m$ are interpreted as cartesian
coordinates on the brane boundary $\partial (AdS_4)$. In
\cite{torinos7} we addressed the question whether such a solvable
parametrization of $AdS_4$ could be extended to a supersolvable
parametrization of anti de Sitter supersapce as defined in
(\ref{ads4N}). In practice that meant to single out a solvable
superalgebra with $4$ bosonic and $4 \times {\cal N}$ fermionic generators.
This turned out to be impossible, yet we easily found a supersolvable
algebra $SSolv_{adS} $ with $4$ bosonic and $2 \times {\cal N}$ fermionic
generators
whose exponential defines {\it solvable anti de Sitter
superspace}:
\begin{equation}
  AdS^{(Solv)}_{4\vert 2{\cal N}}\equiv \exp\left[ SSolv_{adS}\right]
\label{solvsup}
\end{equation}
The supermanifold (\ref{solvsup}) is also a supercoset of the same
supergroup $Osp({\cal N} \vert 4)$ but with respect to a different
subgroup:
\begin{equation}
AdS^{(Solv)}_{4\vert 2{\cal N}} = \frac{Osp(4\vert {\cal N})}
{CSO(1,2\vert{\cal N})}
\label{supcos2}
\end{equation}
where $CSO(1,2\vert {\cal N})\subset Osp({\cal N}\vert 4)$ is
an algebra containing $3 + 3+ \ft{{\cal N}({\cal N}-1)}{2}$ bosonic generators
and $2\times {\cal N}$ fermionic ones. This algebra is the semidirect
product:
\begin{equation}
\begin{array}{ccc}
 CSO(1,2\vert {\cal N}) & = & {\underbrace {ISO(1,2\vert {\cal N})  \times  SO({\cal
 N})}}\\
 \null & \null & \mbox{semidirect}
 \end{array}
\label{CSOdefi}
\end{equation}
of  ${\cal N}$--extended {\it superPoincar\'e} algebra
in $d=3$ ($ISO(1,2\vert {\cal N})$) with the orthogonal group $SO({\cal
N})$.
It should be clearly distinguished from the central extension of the
Poincar\'e superalgebra $Z\left [ISO(1,2\vert {\cal N})\right ]$ which
has the same number of generators but different commutation
relations.
Indeed there are three essential differences that it is worth to
recall at this point:
\begin{enumerate}
  \item In $Z\left [ISO(1,2\vert {\cal N})\right ]$
  the ${{\cal N}({\cal N}-1)}/{2}$ internal generators
  $Z^{ij}$ are abelian, while in $CSO(1,2\vert {\cal N})$ the
  corresponding $T^{ij}$ are non abelian and generate $SO({\cal N})$.
  \item In $Z\left [ISO(1,2\vert {\cal N})\right ]$ the supercharges $q^{\alpha i}$
  commute with $Z^{ij}$ (these are in fact central charges), while in
  $CSO(1,2\vert {\cal N})$ they transform as vectors under $T^{ij}$
  \item In $Z\left [ISO(1,2\vert {\cal N})\right ]$ the anticommutator of two supercharges
  yields, besides the translation generators $P_m$, also the central charges
  $Z^{ij}$, while in $CSO(1,2\vert {\cal N})$ this is not true.
\end{enumerate}
We will see the exact structure of $CSO(1,2\vert {\cal N}) \subset Osp({\cal N} \vert 4)$
and of $ ISO(1,2\vert {\cal N}) \subset CSO(1,2\vert {\cal N})$ as soon as we
have introduced the full orthosymplectic algebra.
In the superconformal interpretation of the $Osp({\cal N}\vert
4)$ superalgebra, $CSO(1,2\vert{\cal N})$ is spanned by the conformal
boosts $K_m$, the Lorentz generators $J^m$ and the special conformal
supersymmetries $s^i_\alpha$. Being a coset, the solvable $AdS$--superspace
$AdS^{(Solv)}_{4\vert 2{\cal N}}$ supports a non linear representation of the full
$Osp({\cal
N}\vert 4 )$ superalgebra. As shown in \cite{torinos7}, we can regard
$AdS^{(Solv)}_{4\vert 2{\cal N}}$ as ordinary anti de Sitter superspace $AdS_{4\vert
{\cal N}}$
where $2\times {\cal N}$ fermionic coordinates have being eliminated by fixing
$\kappa$--supersymmetry.
\par
Our strategy to construct the boundary superfields is the following.
First we construct  the supermultiplets in the bulk by acting on the
abstract states spanning the UIR with the coset
representative of the solvable superspace $AdS^{(Solv)}_{4\vert 2{\cal N}}$
and then we reach the boundary by performing the limit $\rho \to 0$
(see fig. \ref{boufil})
\iffigs
\begin{figure}
\caption{Boundary superfields are obtained as limiting values of superfields in the bulk
\label{boufil}}
\begin{center}
\epsfxsize = 10cm
\epsffile{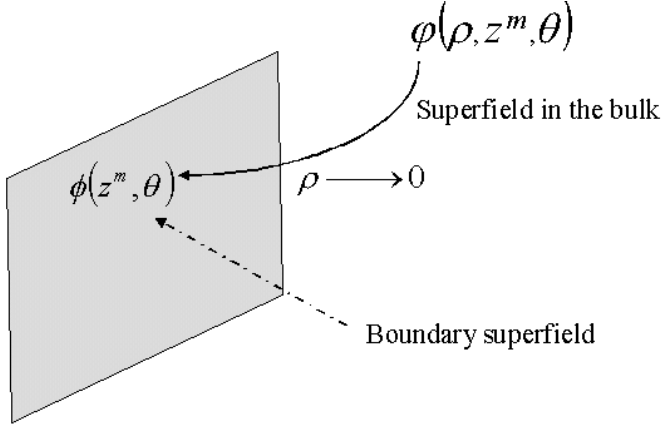}
\vskip  0.2cm
\hskip 2cm
\unitlength=1.1mm
\end{center}
\end{figure}
\fi
\par
\par
The general structure of the $Osp(2\vert 4)$ supermultiplets that may appear in
Kaluza Klein supergravity has been  determined recently in \cite{m111spectrum}
through  consideration of a specific example, that where the
manifold $X^7$ is the sasakian $M^{1,1,1}$. Performing harmonic
analysis on $M^{1,1,1}$ we have found {\it graviton, gravitino} and {\it vector
multiplets}
both in a {\it long} and in a {\it shortened} version. In
addition we have found hypermultiplets that are always short and
the ultra short multiplets corresponding to massless fields.
According to our previous discussion each of these multiplets must
correspond to a primary superfield on the boundary.
We determine such superfields with the above described method.
Short supermultiplets correspond to {\it constrained superfields}.
The shortening conditions  relating mass and hypercharges are
retrieved here as the necessary condition to mantain the constraints
after a superconformal transformation.
\par
As we anticipated above these primary conformal fields are eventually
realized as composite operators in a suitable ${\cal N}=2$, $d=3$ gauge theory.
Hence, in the second part of this paper we construct the general form
of such a theory. To this effect, rather than superspace formalism
we employ our favorite rheonomic approach that, at the end of the
day, yields an explicit component form of the lagrangian and the
supersymmetry transformation rules for all the physical and auxiliary fields.
Although supersymmetric gauge theories in $d=3$ dimensions have been
discussed in the literature through many examples (mostly using superspace
formalism) a survey of their general form seems to us useful.
Keeping track of all the possibilities we construct a supersymmetric off--shell
 lagrangian that employs all the auxiliary fields and
includes, besides minimal gauge couplings and superpotential
interactions also Chern Simons interactions and Fayet Iliopoulos
terms. We restrict however the kinetic terms of the gauge fields to
be quadratic since we are interested in microscopic gauge theories
and not in effective lagrangians. Generalization of our results to
non minimal couplings including arbitrary holomorphic
functions of the scalars in front of the gauge kinetic terms
is certainly possible  but it is not covered in our paper.
\par
In particular we present general formulae for the scalar potential
and we analyse the further conditions that an ${\cal N}=2$ gauge theory
should satisfy in order to admit either ${\cal N}=4$ or ${\cal N}=8$
supersymmetry. This is important in connection with the problem of
deriving the ultraviolet orbifold gauge theories associated with the
sasakian horizons (\ref{sasaki}). Indeed a possible situation that
might be envisaged is that where at the orbifold point the gauge
theory has larger supersymmetry broken to ${\cal N}=2$ by some of
the perturbations responsible for the singularity resolution. It is
therefore vital to write ${\cal N}=4$ and ${\cal N}=8$ theories in
${\cal N}=2$ language. This is precisely what we do here.
%
%
\subsection{Our paper is organized as follows:}
In section \ref{ospn4al} we discuss the definition and the general
properties of the orthosymplectic $Osp({\cal N}\vert 4)$
superalgebra. In particular we discuss its two five--gradings:
compact and non compact, the first related to the supergravity
interpretation , the second to the superconformal field theory
interpretation.
\par
In section \ref{supercoset} we discuss the supercoset structure of
superspace and the realization of the $Osp({\cal N}\vert 4)$
superalgebra as an algebra of transformations in two different
supercosets, the first describing the bulk of $AdS_4$, the second its
boundary $\partial ( AdS_4)$.
\par
In section \ref{supfieldbuetbo} we come to one of the main points
of our paper and focusing on the case ${\cal N}=2$
we show how to construct boundary conformal superfields
out of the Kaluza Klein $Osp(2\vert 4)$ supermultiplets.
\par
In section \ref{n2d3gauge} we discuss the rheonomic construction of a
generic ${\cal N}=2$, $d=3$ gauge theory with arbitrary field content
and arbitrary superpotential interactions.
\par
In section \ref{conclu} we briefly summarize our conclusions.
\par
Finally in appendix \ref{derivationkillings} the reader can find the
explicit derivation of the Killing vectors generating the action of
the $Osp({\cal N}\vert 4)$ superalgebra on superspace. These Killing
vectors are an essential tool for the derivation of our result in
section \ref{supfieldbuetbo}
%
%
\section{The $Osp({\cal N}\vert 4)$ superalgebra: definition, properties and
notations}
\label{ospn4al}
The non compact superalgebra $Osp({\cal N} \vert 4)$ relevant to the
$AdS_4/CFT_3$ correspondence is a real section of the complex orthosymplectic
superalgebra $Osp({\cal N} \vert 4, \IC)$ that admits the Lie algebra
\begin{equation}
  G_{even} = Sp(4,\IR) \times SO({\cal N}, \IR)
\label{geven}
\end{equation}
as even subalgebra. Alternatively, due to the isomorphism
$Sp(4,\IR)\equiv Usp(2,2)$ we can take a different real section of
$Osp({\cal N} \vert 4, \IC)$ such that the even subalgebra is:
\begin{equation}
  G_{even}^\prime = Usp(2,2) \times SO({\cal N}, \IR)
\label{gevenp}
\end{equation}
In this paper we mostly rely on the second formulation (\ref{gevenp})
which is more convenient to discuss unitary irreducible
representations, while in (\cite{torinos7}) we used the first
(\ref{geven}) that is more advantageous for the description of the
supermembrane geometry. The two formulations are related by a unitary
transformation that, in spinor language, corresponds to a different
choice of the gamma matrix representation. Formulation (\ref{geven}) is
obtained in a Majorana representation where all the gamma matrices are real
(or purely imaginary), while formulation (\ref{gevenp}) is related to
a Dirac representation.
\par
Our choice for the gamma matrices in a Dirac representation is the
following one\footnote{we adopt as explicit representation of the $SO(3)~\tau$
matrices a permutation of the canonical Pauli matrices $\s^a$:
$\tau^1=\s^3$, $\tau^2=\s^1$ and $\tau^3=\s^2$;
for the spin covering of $SO(1,2)$ we choose instead the matrices
$\gamma$ defined in (\ref{gammamatrices}).}:
\begin{equation}
\Gamma^0=\left(\begin{array}{cc}
\unity&0\\
0&-\unity
\end{array}\right)\,,\qquad
\Gamma^{1,2,3}=\left(\begin{array}{cc}
0&\tau^{1,2,3}\\
-\tau^{1,2,3}&0
\end{array}\right)\,,
\qquad C_{[4]}=i\Gamma^0\Gamma^3\,,
\label{dirgamma}
\end{equation}
having denoted by $C_{[4]}$ the charge conjugation matrix in $4$--dimensions
$C_{[4]}\, \Gamma^\mu
\, C_{[4]}^{-1} = - ( \Gamma^\mu)^T$.
\par
Then the $Osp({\cal N}\vert 4)$ superalgebra is defined as the set of
graded $(4+{\cal N})\times(4+{\cal N})$ matrices $\mu$ that satisfy the
following two conditions:
\begin{equation}
\begin{array}{rccclcc}
\mu^T & \left( \matrix {C_{[4]} & 0 \cr 0 & \bfone_{{\cal N}\times {\cal N}}
\cr}
\right)& +&\left( \matrix {C_{[4]} & 0 \cr 0 & \bfone_{{\cal N}\times {\cal N}}
\cr}
\right)& \mu & = & 0  \\
\null&\null&\null&\null&\null&\null&\null\\
\mu^\dagger & \left( \matrix {\Gamma^0 & 0 \cr 0 & -\bfone_{{\cal N}\times {\cal
N}} \cr}
\right)& +&\left( \matrix {\Gamma^0 & 0 \cr 0 & -\bfone_{{\cal N}\times {\cal
N}} \cr}
\right)& \mu & = & 0 \\
\end{array}
\label{duecondo}
\end{equation}
the first condition defining the complex orthosymplectic algebra, the
second one the real section with even subalgebra as in
eq.(\ref{gevenp}). Eq.s (\ref{duecondo}) are solved by setting:
\begin{equation}
\mu   =  \left( \matrix {\varepsilon^{AB} \, \frac{1}{4} \,
\left[\IGam_A \, , \, \IGam_B \right ] & \epsilon^i \cr
{\bar \epsilon}^i & \mbox{i}\, \varepsilon_{ij}\, t^{ij} \cr }\right)   \nonumber\\
\label{mumatri}
\end{equation}
In eq.({\ref{mumatri}) $\varepsilon_{ij}=-\varepsilon_{ji}$
is an arbitrary real antisymmetric ${\cal N} \times {\cal N}$ tensor,
 $t^{ij} = -t^{ji}$ is the antisymmetric ${\cal N} \times {\cal N}$
 matrix:
\begin{equation}
 ( t^{ij})_{\ell m} = \mbox{i}\left( \delta ^i_\ell \delta ^j_m - \delta
 ^i_m \delta ^j_\ell \right)
\label{tgene}
\end{equation}
namely a standard generator of the $SO({\cal N})$ Lie algebra,
\begin{equation}
  \IGam_A=\cases{\begin{array}{cl}
  \mbox{i} \, \Gamma_5 \Gamma_\mu & A=\mu=0,1,2,3 \\
\Gamma_5\equiv\mbox{i}\Gamma^0\Gamma^1\Gamma^2\Gamma^3 & A=4 \\
\end{array} }
\label{bigamma}
\end{equation}
denotes a realization of the $SO(2,3)$ Clifford algebra:
\begin{eqnarray}
  \left \{ \IGam_A \, ,\, \IGam_B \right\} &=& 2
  \eta_{AB}\nonumber\\
\eta_{AB}&=&{\rm diag}(+,-,-,-,+)
\label{so23cli}
\end{eqnarray}
and
\begin{equation}
  \epsilon^i = C_{[4]} \left( {\bar \epsilon}^i\right)^T
  \quad (i=1,\dots\,{\cal N})
\label{qgene}
\end{equation}
are ${\cal N}$ anticommuting Majorana spinors.
\par
The index conventions we have so far introduced can be
summarized as follows. Capital indices $A,B=0,1,\ldots,4$ denote
$SO(2,3)$ vectors. The   latin indices of type
$i,j,k=1,\ldots,{\cal N}$ are $SO({\cal N})$ vector indices.
The indices $a,b,c,\ldots=1,2,3$ are used to denote spatial directions
of $AdS_4$: $\eta_{ab}={\rm diag}(-,-,-)$, while the indices of type
$m,n,p,\ldots=0,1,2$ are space-time indices for the Minkowskian
boundary $\partial \left( AdS_4\right) $: $\eta_{mn}={\rm diag}(+,-,-)$.
To write the $Osp({\cal N} \vert 4)$ algebra in abstract form it
suffices to read the graded matrix (\ref{mumatri}) as a linear
combination of generators:
\begin{equation}
  \mu \equiv -\mbox{i}\varepsilon^{AB}\, M_{AB}
  +\mbox{i}\varepsilon_{ij}\,T^{ij}
  +{\bar \epsilon}_i \, Q^i
\label{idegene}
\end{equation}
where $Q^i = C_{[4]} \left(\overline Q^i\right)^T$ are also Majorana spinor
operators.
Then the superalgebra reads as follows:
\begin{eqnarray}
\left[ M_{AB} \, , \, M_{CD} \right]  & = & \mbox{i} \, \left(\eta_{AD} M_{BC}
+ \eta_{BC} M_{AD} -\eta_{AC} M_{BD} -\eta_{BD} M_{AC} \right)  \nonumber\\
\left [T^{ij} \, , \, T^{kl}\right] &=&
-\mbox{i}\,(\delta^{jk}\,T^{il}-\delta^{ik}\,T^{jl}-
\delta^{jl}\,T^{ik}+\delta^{il}\,T^{jk})\,  \nonumber \\
\left[M_{AB} \, , \, Q^i \right] & = & -\mbox{i} \frac{1}{4} \,
\left[\IGam_A \, , \, \IGam_B \right ] \, Q^i \nonumber\\
\left[T^{ij}\, , \,  Q^k\right] &=&
-\mbox{i}\, (\delta^{jk}\, Q^i - \delta^{ik}\,   Q^j ) \nonumber\\
\left \{Q^{\alpha i}, \overline Q_{\b}^{\,j}\right \} & = & \mbox{i}
\delta^{ij}\frac{1}{4}\,\left[\IGam^A \, , \, \IGam^B \right ]{}^{\a}_{\
\b}M_{AB}
+\mbox{i}\delta^{\a}_{\,\b}\,T^{ij}
\label{pippa}
\end{eqnarray}
The form (\ref{pippa}) of the $Osp({\cal N}\vert 4)$ superalgebra
coincides with that given in papers \cite{freedmannicolai},\cite{multanna} and
utilized by us in our recent derivation of the $M^{111}$ spectrum
\cite{m111spectrum}.
\par
In the gamma matrix basis (\ref{dirgamma})
the Majorana supersymmetry charges have the following form:
\begin{eqnarray}
Q^i =
\left(\matrix{a_\a^i\cr\varepsilon_{\alpha\beta}\bar a^{\b i}}\right)\,,
\qquad \bar a^{\alpha i} \equiv \left( a_\alpha^i \right)^\dagger \,,
\end{eqnarray}
where $a_\a^i$ are two-component $SL(2,\IC)$ spinors: $\a,\b,\ldots = 1,2$.
We do not use dotted and undotted indices to
denote conjugate $SL(2,\IC)$ representations; we rather use higher and
lower indices.
Raising and lowering  is performed by means of the
$\varepsilon$-symbol:
\begin{equation}
\psi_\a = \varepsilon_{\a\b} \psi^\b \,, \qquad
\psi^\a = \varepsilon^{\a\b} \psi_\b\,,
\end{equation}
where $\varepsilon_{12}=\varepsilon^{21}=1$, so that
$\varepsilon_{\a\g}\varepsilon^{\g\b}=\delta_\a^\b$.
Unwritten indeces are contracted according to the rule ``{\it from eight to
two}''.
\par
In the second part of the paper where we deal with the
lagrangian of $d=3$ gauge theories, the conventions for  two--component
spinors are slightly modified to simplify the notations and avoide the explicit
writing of spinor indices.
The Grassman coordinates of  ${\cal N}\!\!=\!\!2$ three-dimensional
superspace introduced in equation (\ref{complexthet}) , $\theta^\pm_\a$, are
renamed $\theta$ and $\theta^c$.
The reason for the superscript ``$\,^c\,$'' is that, in three
dimensions the upper and lower components of the four--dimensional
$4$--component spinor are charge conjugate:
\begin{equation}\label{conjugations}
\theta^c\equiv C_{[3]}\overline\theta^T\,,\qquad
\overline\theta\equiv\theta^\dagger\g^0\,,
\end{equation}
where $C_{[3]}$ is the $d=3$ charge conjugation matrix:
\begin{equation}
\left\{\begin{array}{ccc}
C_{[3]}\g^m C_{[3]}^{-1}&=&-(\g^m)^T\\
\g^0\g^m(\g^0)^{-1}&=&(\g^m)^\dagger
\end{array}\right.
\end{equation}
The lower case gamma matrices are $2\!\times\!2$
and provide a realization of the $d\!=\!2\!+\!1$ Clifford algebra:
\begin{equation}
  \{\gamma^m \, , \, \gamma^n \} = \eta^{mn}
\label{so21cli}
\end{equation}
Utilizing  the following explicit basis:
\begin{equation}\label{gammamatrices}
\left\{\begin{array}{ccl}
\g^0&=&\sigma^2\\
\g^1&=&-i\sigma^3\\
\g^2&=&-i\sigma^1
\end{array}\right.\qquad C_{[3]}=-i\sigma^2\,,
\end{equation}
both $\g^0$ and $C_{[3]}$ become proportional to $\varepsilon_{\a\b}$.
This implies that in equation (\ref{conjugations}) the role of the
matrices $C_{[3]}$ and $\g^0$ is just to convert  upper  into lower
$SL(2,\IC)$ indices and viceversa.
\par
The relation between the two notations for the spinors is summarized
in the following table:
\begin{equation}
\begin{array}{|c|c|}
\hline
(\theta^+)^\a&\sqrt{2}\,\theta\\
(\theta^+)_\a&\sqrt{2}\,\overline\theta^c\\
(\theta^-)^\a&-i\sqrt{2}\,\theta^c\\
(\theta^-)_\a&-i\sqrt{2}\,\overline\theta\\
\hline
\end{array}
\end{equation}
With the second set of conventions the spinor indices can
be ignored since the contractions are always made between
barred (on the left) and unbarred (on the right) spinors,
corresponding to the ``{\it eight to two}'' rule of the first set of
conventions.
Some examples of this ``{\it transcription}'' are given by:
\begin{eqnarray}
\overline\theta\theta=\ft{1}{2}i(\theta^-)_\a(\theta^+)^\a\nonumber\\
\overline\theta^c\g\theta^c=\ft{1}{2}i(\theta^+)_\a(\g)^\a_{\ \b}
(\theta^-)^\b
\end{eqnarray}
%
%
\subsection{Compact and non compact five gradings of the $Osp({\cal N}|4)$
superalgebra}
As it is extensively explained in \cite{gunaydinminiczagerman1}, a non-compact
group $G$  admits unitary irreducible representations of the lowest
weight type if it has a maximal compact subgroup $G^0$ of the form
$G^0=H\times U(1)$ with respect to whose Lie algebra $g^0$ there
exists has a {\it three grading} of the Lie algebra $g$ of $G$.
In the case of a non--compact superalgebra the lowest weight
UIR.s can be constructed if the three grading is   generalized to
a {\it five grading} where the even (odd) elements are integer
(half-integer) graded:
\begin{eqnarray}
g = g^{-1} \oplus g^{-\ft12} \oplus g^0 \oplus g^{+\ft12} \oplus g^{+1}\,,\\
\nonumber\\
\left[g^k,g^l\right]\subset g^{k+l}\qquad g^{k+l}=0\ {\rm for}\ |k+l|>1\,.
\end{eqnarray}
For the supergroup
$Osp({\cal N}|4)$) this grading can be made in two ways, choosing
as grade zero subalgebra either the maximal compact subalgebra
\begin{eqnarray}
g^0 \equiv SO(3) \times SO(2) \times SO({\cal N}) \subset Osp({\cal N} \vert 4)
\label{so3so2}
\end{eqnarray}
or the non-compact subalgebra
\begin{eqnarray}
{\tilde g}^0 \equiv
SO(1,2) \times SO(1,1) \times SO({\cal N}) \subset Osp({\cal N} \vert 4)
\label{so12so11}
\end{eqnarray}
which also exists, has the same complex extension and is also
maximal.
\par
The existence of the double five--grading is the algebraic core of
the $AdS_4/CFT_3$ correspondence. Decomposing a UIR  of
$Osp({\cal N} \vert 4)$ into representations of $g^0$ shows its
interpretation as a supermultiplet of {\it particles states} in the bulk of
$AdS_4$, while decomposing it into representations of ${\tilde g}^0$
shows its interpretation as a supermultiplet of {\it conformal
primary fields} on the boundary $\partial (AdS_4)$.
\par
In both cases the grading is determined by the generator $X$ of the abelian
factor $SO(2)$ or $SO(1,1)$:
\begin{equation}
[X,g^k]=k\,g^k
\end{equation}
In the compact case (see \cite{freedmannicolai}) the $SO(2)$ generator
$X$ is given by $M_{04}$.
It is interpreted as the energy generator of the four-dimensional $AdS$
theory.
It was used in \cite{multanna} and \cite{m111spectrum} for the construction
of the $Osp(2 \vert 4)$ representations, yielding the long
multiplets of \cite{multanna} and the short and ultra-short
multiplets of \cite{m111spectrum}.
We repeat such decompositions here.
\par
We call $E$ the energy generator of $SO(2)$, $L_a$ the rotations
of $SO(3)$:
\begin{eqnarray}
E &=& M_{04} \,, \nonumber \\
L_a &=& \ft12 \varepsilon_{abc} \, M_{bc} \,,
\end{eqnarray}
and $M_a^\pm$ the boosts:
\begin{eqnarray}
M_a^+ &=& - M_{a4} + i M_{0a} \,, \nonumber \\
M_a^- &=&  M_{a4} + i M_{0a} \,.
\end{eqnarray}
The supersymmery generators are $a^i_\alpha$ and $\bar a^{\alpha i}$.
Rewriting the $Osp({\cal N} \vert 4)$ superalgebra (\ref{pippa})
in this basis we obtain:
\begin{eqnarray}
{}[E, M_a^+] &=& M_a^+ \,,\nonumber \\
{}[E, M_a^-] &=& -M_a^- \,,\nonumber \\
{}[L_a, L_b] &=& i \, \varepsilon_{abc} L_c
\,, \nonumber \\
{}[M^+_a, M^-_b] &=& 2 \, \delta_{ab}\, E + 2 i \,
\varepsilon_{abc} \, L_c \,,
\nonumber \\
{}[L_a, M^+_b ] &=& i \, \varepsilon_{abc} \, M^+_c \,,
\nonumber \\
{}[L_a, M^-_b ] &=& i \, \varepsilon_{abc} \, M^-_c \,,
\nonumber \\
{}[T^{ij}, T^{kl}] &=&
-i\,(\delta^{jk}\,T^{il}-\delta^{ik}\,T^{jl}-
\delta^{jl}\,T^{ik}+\delta^{il}\,T^{jk})
\,, \nonumber \\
{}[T^{ij}, \bar a^{\alpha k}] &=&
-i\, (\delta^{jk}\, \bar a^{\alpha i} - \delta^{ik}\, \bar a^{\alpha j} )
\,, \nonumber \\
{}[T^{ij},  a_\alpha^k] &=&
-i\, (\delta^{jk}\, a_\alpha^i - \delta^{ik}\,   a_\alpha^i )
\,,\nonumber \\
{}[E, a_\alpha^i] &=& -\ft12 \, a_\alpha^i \,, \nonumber \\
{}[E, \bar a^{\alpha i}] &=& \ft12 \, \bar a^{\alpha i}
\,, \nonumber \\
{}[M_a^+, a_\alpha^i] &=& (\tau_a)_{\alpha\beta}\, \bar a^{\beta i}
\,, \nonumber \\
{}[M_a^-, \bar a^{\alpha i}] &=&
- (\tau_a)^{\alpha\beta} \,  a_\beta^i \,,\nonumber \\
{}[L_a, a_\alpha^i]&=& \ft12 \, (\tau_a)_{\alpha}{}^\beta \,  a^i_\beta
\,, \nonumber \\
{}[L_a, \bar a^{\alpha i}] &=& -\ft12 \, (\tau_a)^\alpha{}_\beta
\,  \bar a^{\beta i} \,, \nonumber \\
\{a_\alpha^i, a_\beta^j \} &=& \delta^{ij} \, (\tau^k)_{\alpha\beta}\,
M_k^- \,, \nonumber \\
\{\bar a^{\alpha i}, \bar a^{\beta j} \} &=&
\delta^{ij} (\tau^k)^{\alpha\beta} \, M_k^+ \,,
\nonumber \\
\{ a_\alpha^i, \bar a^{\beta j} \}
&=& \delta^{ij}\, \delta_\alpha{}^\beta \, E
+ \delta^{ij} \, (\tau^k)_\alpha{}^\beta \, L_k
+i \, \delta_\alpha{}^\beta \, T^{ij} \,.
\label{ospE}
\end{eqnarray}
The five--grading structure of the algebra (\ref{ospE}) is shown in
fig. \ref{pistac}
In the superconformal field theory context we are interested in the
action of the $Osp({\cal N} \vert 4)$
generators on superfields living on the minkowskian boundary  $\partial(AdS_4)$.
To be precise the boundary   is a compactification of $d=3$
Minkowski space and admits a conformal family of metrics
$g_{mn} = \phi(z) \eta_{mn}$ conformally
equivalent to the the flat Minkowski metric
\begin{equation}
\eta_{mn} = (+,-,-) \,, \qquad m,n,p,q = 0,1,2 \,.
\label{minkio3}
\end{equation}
Precisely because we are interested in conformal field theories the
the choice of representative metric inside the conformal
family is immaterial and the flat one (\ref{minkio3}) is certainly the
most convenient.
The requested action  of the superalgebra generators
is obtained upon starting from  the non--compact
grading with respect to (\ref{so12so11}). To this effect we define
the {\it dilatation} $SO(1,1)$ generator $D$ and the {\it Lorentz} $SO(1,2)$
generators $J^m$ as follows:
\begin{equation}
D \equiv i\, M_{34} \,, \qquad J^m = \ft{i}{2} \, \varepsilon^{mpq} M_{pq}\,.
\label{dilalor}
\end{equation}
\begin{figure}[ht]
\begin{center}
\leavevmode
\hbox{%
\epsfxsize=11cm
\epsfbox{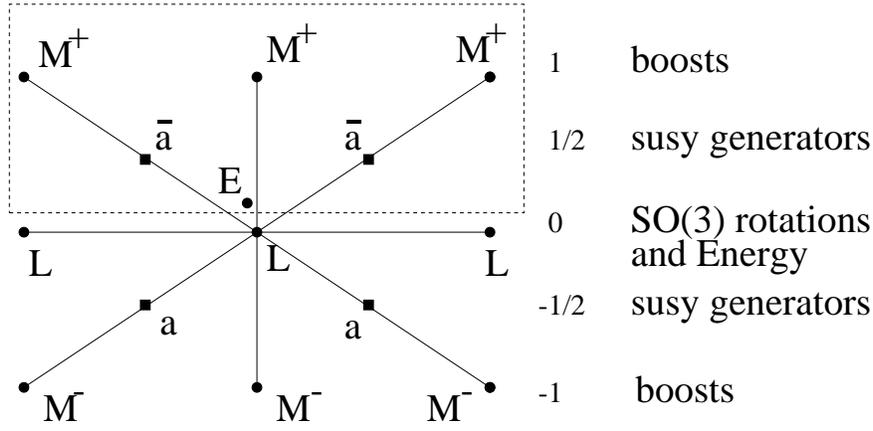}}
\caption{{\small Schematic representation of the root diagram
of $Osp({\cal N}|4)$ in the $SO(2) \times SO(3)$ basis.
\label{pistac}
The grading w.r.t. the energy $E$ is given on the right. }}
\end{center}
\end{figure}
In addition we define the
the $d=3$ {\it translation generators} $P_m$  and {\it special conformal
boosts} $K_m$ as follows:
\begin{eqnarray}
P_m = M_{m4} - M_{3m} \,, \nonumber \\
K_m = M_{m4} + M_{3m} \,.
\label{Pkdefi}
\end{eqnarray}
Finally we define the generators of $d=3$ {\it ordinary} and {\it special
conformal supersymmetries}, respectively given by:
\begin{eqnarray}
q^{\alpha i} = \ft{1}{\sqrt{2}}\left(a_\alpha^i + \bar a^{\alpha i}\right)
\,, \nonumber \\
s_\alpha^i = \ft{1}{\sqrt{2}}\left(-a_\alpha^i + \bar a^{\alpha i}\right)
\,.
\label{qsdefi}
\end{eqnarray}
The $SO({\cal N})$ generators are left unmodified as above.
In this new basis the $Osp({\cal N}\vert 4)$-algebra (\ref{pippa}) reads
as follows
\begin{eqnarray}
{}[D, P_m] &=& -P_m \,, \nonumber \\
{}[D, K_m] &=& K_m \,, \nonumber \\
{}[J_m, J_n] &=& \varepsilon_{mnp} \, J^p \,, \nonumber \\
{}[K_m, P_n] &=& 2 \, \eta_{mn}\, D - 2 \, \varepsilon_{mnp} \, J^p \,,
\nonumber \\
{}[J_m, P_n] &=&  \varepsilon_{mnp} \, P^p \,, \nonumber \\
{}[J_m, K_n] &=&  \varepsilon_{mnp} \, K^p \,, \nonumber \\
{}[T^{ij}, T^{kl}] &=&
-i\,(\delta^{jk}\,T^{il}-\delta^{ik}\,T^{jl}-
\delta^{jl}\,T^{ik}+\delta^{il}\,T^{jk})
\,, \nonumber \\
{}[T^{ij}, q^{\alpha k}] &=&
-i\, (\delta^{jk}\, q^{\alpha i} - \delta^{ik}\, q^{\alpha j} )
\,, \nonumber \\
{}[T^{ij}, s_\alpha^k] &=&
-i\, (\delta^{jk}\, s_\alpha^i - \delta^{ik}\,  s_\alpha^i )
\,,\nonumber \\
{}[D, q^{\alpha i}] &=& -\ft12 \, q^{\alpha i} \,,\nonumber \\
{}[D, s_\alpha^i] &=& \ft12 \, s_\alpha^i \,,\nonumber \\
{}[K^m, q^{\alpha i} ] &=&
- i\, (\gamma^m)^{\alpha \beta}\, s_\beta^i \,, \nonumber
\\
{}[P^m, s_\alpha^i] &=&
- i\, (\gamma^m)_{\alpha \beta}\, q^{\beta i} \,, \nonumber \\
{}[J^m, q^{\alpha i} ] &=&
- \ft{i}{2} \, (\gamma^m)^\alpha{}_\beta q^{\beta i} \,,
\nonumber \\
{}[J^m, s_\alpha^i] &=& \ft{i}{2} \,  (\gamma^m)_\alpha{}^\beta s_\beta^i \,,
\nonumber \\
\{q^{\alpha i}, q^{\beta j} \} &=& - i\, \delta^{ij}\,
(\gamma^m)^{\alpha\beta} P_m \,, \nonumber \\
\{s_\alpha^i, s_\beta^j \} &=&
i\, \delta^{ij} \, (\gamma^m)_{\alpha\beta} K_m \,,
\nonumber \\
\{q^{\alpha i}, s_\beta^j \} &=& \delta^{ij} \delta^\alpha{}_\beta \, D
- i\, \delta^{ij} (\gamma^m)^\alpha{}_\beta J_m
+ i \delta^\alpha{}_\beta T^{ij}
\,.
\label{ospD}
\end{eqnarray}
and the five grading structure of eq.s (\ref{ospD}) is displayed in
fig.\ref{pirillo}.
\begin{figure}[ht]
\begin{center}
\leavevmode
\hbox{%
\epsfxsize=11cm
\epsfbox{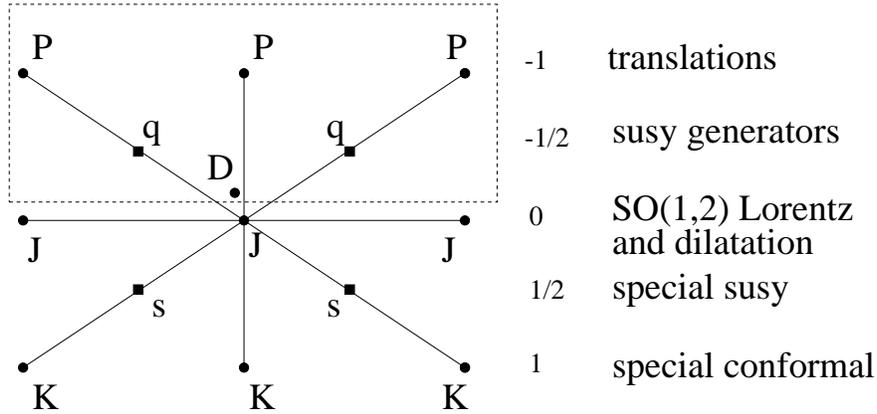}}
\caption{{\small Schematic  representation of the root diagram
of $Osp({\cal N}|4)$ in the $SO(1,1)\times SO(1,2)$ basis.
The grading w.r.t. the dilatation $D$ is given on the right.\label{pirillo}}}
\end{center}
\end{figure}
In both cases of fig.\ref{pistac} and fig.\ref{pirillo} if one takes
the subset of generators of positive grading plus the abelian grading
generator $X=\cases{E\cr D\cr}$ one obtains a {\it solvable superalgebra} of dimension
$4+2{\cal N}$. It is however only in the non compact case of
fig.\ref{pirillo} that the bosonic subalgebra of the solvable
superalgebra generates anti de Sitter space $AdS_4$ as a solvable group
manifold. Therefore the solvable superalgebra $Ssolv_{adS}$ mentioned
in eq. (\ref{solvsup}) is the vector span of the following
generators:
\begin{equation}
  Ssolv_{adS} \equiv \mbox{span} \left\{ P_m, D, q^{\alpha i} \right\}
\label{Span}
\end{equation}
%
%
\subsection{The lowest weight UIR.s as seen from the compact and non
compact five--grading viewpoint}
The structure of all the $Osp(2\vert 4)$ supermultiplets
relevant to Kaluza Klein supergravity is known. Their spin content is upper
bounded by
$s=2$ and they fall into three classes: {\it long, short} and {\it ultrashort}.
Such a result has  been obtained in \cite{m111spectrum} by explicit harmonic
analysis on $X^7 = M^{111}$, namely through the analysis of a specific
example of ${\cal N}\!\!=\!\!2$ compactification on $AdS_4 \times X^7$.
As stressed in the introduction  the goal of the present paper
is to reformulate the structure of these multiplets in  a way
appropriate for comparison  with {\it composite operators} of the
 three-dimensional gauge theory living on the boundary
 $\partial(AdS_4)$ that behave as {\it primary conformal fields}.
Actually, in view of the forthcoming
Kaluza-Klein spectrum on $X^7=N^{010}$ \cite{n010},
that is arranged into $Osp(3\vert 4)$ rather than $Osp(2 \vert 4)$ multiplets,
it is more convenient to begin by discussing
$Osp({\cal N}\vert 4)$ for generic ${\cal N}$.
\par
We start by briefly
recalling the procedure of \cite{freedmannicolai,heidenreich}
to construct UIR.s of $Osp({\cal N}\vert 4)$ in the compact grading
(\ref{so3so2}). Then, in a parallel way to
what was done in \cite{gunaydinminiczagerman2} for the case
of the $SU(2,2\vert 4)$ superalgebra we show that also
for $Osp({\cal N}\vert 4)$in each UIR carrier space  there exists a
unitary rotation
that maps  eigenstates of $E, L^2, L_3$
into eigenstates of $D,J^2,J_2$. By means of such a rotation the
decomposition of the UIR into $SO(2)\times SO(3)$ representations is
mapped into an analogous decomposition into $SO(1,1) \times SO(1,2)$
representations.
While $SO(2)\times SO(3)$ representations describe the {\it on--shell}
degrees of freedom of a {\it bulk particle} with an energy
$E_0$ and a spin $s$,  irreducible  representations of $SO(1,1) \times
SO(1,2)$ describe the {\it off-shell} degrees of freedom  of a
{\it boundary field} with scaling weight $D$ and Lorentz character $J$.
Relying on this  we show how to
construct the on-shell four-dimensional
superfield multiplets that generate the states of these representations
and the off-shell three-dimensional superfield multiplets that build
the conformal field theory on the boundary.
\par
Lowest weight representations of $Osp({\cal N}\vert 4)$  are
constructed starting from the basis (\ref{ospE}) and choosing a
{\it a vacuum state} such that
\begin{eqnarray}
M_i^- \vert (E_0, s, \Lambda) \rangle &=& 0 \,,
\nonumber \\
a^i_\alpha \vert (E_0, s, \Lambda) \rangle &=& 0 \,,
\label{energyreps}
\end{eqnarray}
where $E_0$ denotes the eigenvalue of the energy operator
$M_{04}$ while $s$ and $\Lambda$ are the labels of an irreducible $SO(3)$ and
$SO({\cal N})$ representation, respectively. In particular we have:
\begin{eqnarray}
M_{04}\, \vert (E_0, s, \Lambda)\rangle
& = & E_0 \, \vert (E_0, s, \Lambda) \rangle
\nonumber \\
L^a \, L^a \, \vert (E_0, s, \Lambda) \rangle & = & s(s+1) \,
\vert (E_0, s, \Lambda) \rangle  \nonumber\\
L^3  \vert (E_0, s, \Lambda) \rangle & =& s \,\vert (E_0, s, \Lambda) \rangle\,.
\label{eigval}
\end{eqnarray}
The states filling up the UIR
are then built by applying the operators $M^-$ and the anti-symmetrized
products of the operators $\bar a^i_\alpha$:
\begin{eqnarray}
\left( M_1^+ \right)^{n_1} \left( M_2^+ \right)^{n_2}
\left( M_3^+ \right)^{n_3} [ \bar a^{i_1}_{\alpha_1}
\dots \bar a^{i_p}_{\alpha_p}]
\vert (E_0, s, \Lambda) \rangle
\label{so3so2states}
\end{eqnarray}
\par
Lowest weight representations are similarly constructed
with respect to five--grading (\ref{ospD}).
One starts from a vacuum state that is annihilated by the conformal boosts and
by
the special conformal supersymmetries
\begin{eqnarray}
K_m \, \vert (D_0, j, \Lambda) \rangle &=& 0 \,,
\nonumber \\
s^i_\alpha \, \vert (D_0, j, \Lambda) \rangle &=& 0 \,,
\label{primstate}
\end{eqnarray}
and that is an eigenstate of the dilatation operator $D$ and an
irreducible $SO(1,2)$ representation of spin $j$:
\begin{eqnarray}
  D \, \vert (D_0, j, \Lambda) \rangle &=& D_0 \, \vert (D_0, j, \Lambda) \rangle
  \nonumber\\
  J^m \, J^n \, \eta_{mn} \,\vert (D_0, j, \Lambda) \rangle &=&
 j (j +1) \,\vert (D_0, j, \Lambda) \rangle\nonumber\\
  J_2 \,\vert (D_0, j, \Lambda) \rangle & = & j \vert (D_0, j, \Lambda) \rangle
\label{Djvac}
\end{eqnarray}
As for the $SO({\cal N})$ representation the new vacuum is the same
as before.
The states filling the UIR are now constructed by applying to the vacuum
the operators $P_m$ and the
anti-symmetrized products of $q^{\alpha i}$,
\begin{eqnarray}
\left( P_0\right)^{p_0} \left( P_1 \right)^{p_1}
\left( P_2 \right)^{p_2} [ q^{\alpha_1 i_1} \dots q^{\alpha_q i_q}]
\vert (D_0, j, \Lambda) \rangle\,.
\label{so12so11states}
\end{eqnarray}
\par
In the language of  conformal field theories the vacuum state
satisfying eq.(\ref{primstate}) is named a {\it primary state}
(corresponding to the value at $z^m=0$ of a primary conformal field}.
The states (\ref{so12so11states}) are called the {\it descendants}.
\par
\begin{figure}[ht]
\begin{center}
\leavevmode
\hbox{%
\epsfxsize=12cm
\epsfbox{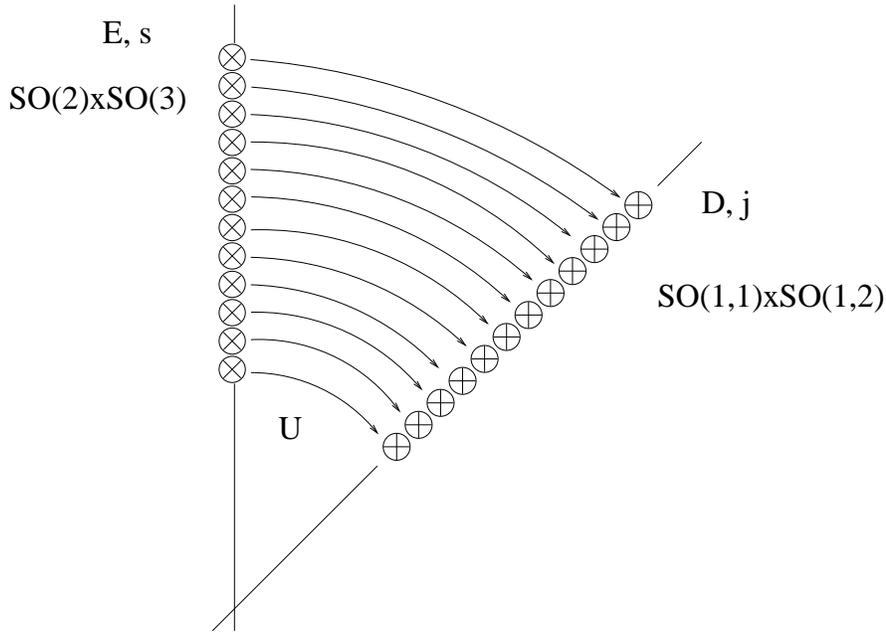}}
\caption{{\small The operator $U=\exp\{i\pi/\sqrt 2(E-D)\}$ rotates
the Hilbert space of the physical states.
It takes states labeled by the Casimirs ($E,\,s$) of
the $SO(2)\times SO(3)\subset Osp({\cal N}|4)$ into states
labeled by the Casimirs ($D,\,j$) of $SO(1,1)\times SO(1,2)$.\label{rota}}}
\end{center}
\end{figure}

The rotation between the $SO(3)\times SO(2)$ basis
and the $SO(1,2)\times SO(1,1)$ basis is performed by the
operator:
\begin{eqnarray}
U\equiv  \exp \left[{\ft{i}{\sqrt{2}}\pi(E-D)} \right] \,,
\label{rotationmatrix}
\end{eqnarray}
which has the following properties,
\begin{eqnarray}
D U &=& - U E \,, \nonumber \\
J_0 U &=& i \, U L_3 \,, \nonumber \\
J_1 U &=& U L_1 \,, \nonumber \\
J_2 U &=& U L_2 \,,
\label{L0}
\end{eqnarray}
with respect to the grade $0$ generators. Furthermore, with respect
to the non vanishing grade generators we have:
\begin{eqnarray}
K_0 U &=& -i \, U M_3^- \,, \nonumber \\
K_1 U &=& - U M_1^- \,, \nonumber \\
K_2 U &=& - U M_2^- \,, \nonumber \\
P_0 U &=& i\,U M_3^+ \,, \nonumber \\
P_1 U &=& U M_1^+ \,, \nonumber \\
P_2 U &=& U M_2^+ \,, \nonumber \\
q^{\alpha i} U &=& -i\, U \bar a^{\alpha i} \nonumber \\
s_\alpha^i U &=& i \, U a_\alpha^i \,.
\label{L+L-}
\end{eqnarray}
As one  immediately sees from (\ref{L+L-}), U interchanges the
compact five--grading structure of the superalgebra with its non
compact one. In particular
the $SO(3)\times SO(2)$-vacuum with energy $E_0$
is mapped into an $SO(1,2)\times SO(1,1)$ primary state and
one obtains all the descendants (\ref{so12so11states}) by acting
with $U$ on the particle states (\ref{so3so2states}). Furthermore
from (\ref{L0}) we read the conformal weight and the Lorentz
group representation of the primary state
$U \vert (E_0, s, J) \rangle$. Indeed its
 eigenvalue with respect to the dilatation generator $D$ is:
\begin{equation}
D_0 = - E_0 \,.
\end{equation}
and  we find the following relation between  the Casimir operators
 of $SO(1,2)$ and $SO(3)$,
\begin{equation}
J^2 U = U L^2 \,, \qquad J^2 \equiv -J_0^2 + J_1^2 + J_2^2 \,,
\end{equation}
which implies that
\begin{equation}
j = s \,.
\end{equation}
Hence under the action of $U$ a particle state of energy $E_0$ and
spin $s$ of the bulk is mapped into a {\it primary conformal field}
of conformal weight $-E_0$ and Lorentz spin $s$ on the boundary.
This discussion is visualized in fig.\ref{rota}
%
%
%
%
%
\section{$AdS_4$ and $\partial AdS_4$ as cosets and their Killing vectors}
\label{supercoset}
In the previous section we  studied $Osp({\cal N}\vert 4)$ and its
representations in two different bases. The form (\ref{ospE}) of the
superalgebra
is  that we used in \cite{m111spectrum} to construct the $Osp(2\vert
4)$ supermultiplets from Kaluza Klein supergravity. It will be similarly used
to obtain the $Osp(3\vert 4)$ spectrum on $X^7=N^{010}$.
We translated these results
in terms of the form (\ref{ospD}) of the $Osp({\cal N}\vert 4)$ algebra
in order to allow a comparison with the three-dimensional
CFT on the boundary.
In this section we introduce the announced description of the anti de
Sitter superspace and of  its boundary in terms of supersolvable Lie algebra
parametrization as in eq.s(\ref{solvsup}),(\ref{supcos2}).
It turns out that such a description is the most appropriate
for a comparative study between $AdS_4$ and its boundary.  We calculate
the Killing vectors of these two coset spaces
since they are needed to determine the
superfield multiplets living on both $AdS_4$ and $\partial AdS_4$.
\par
So we write both the bulk and the boundary superspaces as
supercosets\footnote{For an
extensive explanation about supercosets we refer the reader to
\cite{castdauriafre}. In the context of $D=11$ and $D=10$ compactifications
see also \cite{renatpiet}},
\begin{eqnarray}
\frac{G}{H}  \,,
\label{GoverH}
\end{eqnarray}
Applying supergroup elements  $g \in Osp({\cal N}\vert 4)$ to the coset
representatives
$L(y)$ these latter transform as follows:
\begin{eqnarray}
g \, L(y) = L(y^\prime) h(g,y) \,,
\label{defcoset}
\end{eqnarray}
where  $h(y)$ is some element of $H \subset Osp({\cal N}\vert 4)$, named
the compensator  that, generically depends both on $g$ and on the coset point
$y\in G/H$.
For our purposes it is useful to consider the infinitesimal
form of (\ref{defcoset}), i.e. for infinitesimal $g$ we can write:
\begin{eqnarray}
g&=& 1 + \epsilon^A T_A \,, \nonumber \\
h &=& 1 + \epsilon^A W^H_A(y) T_H \,, \nonumber \\
y^{\mu\prime}  \, & = & y^\mu+\epsilon^A k^\mu_A(y)
\end{eqnarray}
and we obtain:
\begin{eqnarray}
T_A L(y) &=& k_A L(y) + L(y) T_H W^H_A(y) \,
,\label{infinitesimalcosettransfo}\\
k_A & \equiv & k^\mu_A(y) \, \frac{\partial}{\partial y^\mu}
\label{kildefi}
\end{eqnarray}
The shifts in the superspace coordinates $y$ determined by the supergroup elements
(see eq.(\ref{defcoset})) define the Killing vector fields
(\ref{kildefi}) of the coset manifold.\footnote{The Killing vectors satisfy the
algebra with structure functions with opposite sign, see \cite{castdauriafre}}
\par
Let us now consider the solvable anti de Sitter superspace
defined in eq.s (\ref{solvsup}),(\ref{supcos2}). It describes a
$\kappa$--gauge fixed supersymmetric extension of the bulk  $AdS_4$.
As explained by eq.(\ref{supcos2}) it is a supercoset
(\ref{GoverH})
where $G=Osp({\cal N}\vert 4)$ and $H=CSO(1,2\vert{\cal N}) \times SO({\cal N})$
 Using the non--compact basis (\ref{ospD}),
the subgroup $H$ is given by,
\begin{eqnarray}
H^{AdS} =CSO(1,2\vert{\cal N}) \,\equiv\,
 \mbox{span}\, \left\{ \, J^m, K_m, s_\alpha^{i}, T^{ij} \,\right\}  \,.
\label{HAdS}
\end{eqnarray}
A coset representative can be written as follows\footnote{We use the notation
$x\cdot y \equiv x^m y_m$ and $\theta^i q^i \equiv \theta^i_\alpha q^{\alpha
i}$. }:
\begin{eqnarray}
L^{AdS}(y) = \exp \left [{\rho D + i \, x \cdot P + \theta^i q^i} \right ]\,,
\qquad
y=(\rho, x,\theta) \,.
\label{AdScosetrepresentative}
\end{eqnarray}
In $AdS_{4\vert 2{\cal N}}$
 $s$-supersymmetry and  $K$-symmetry have a non linear realization
 since the corresponding generators are not part of the solvable superalgebra
$Ssolv_{adS}$  that is exponentiated (see eq.(\ref{Span}).
\par
The form of the Killing vectors simplifies considerably if we
rewrite the coset representative as a product of exponentials
\begin{eqnarray}
L(y) = \exp \left[ i \, z \cdot P\right]\, \cdot \, \exp \left[ \xi^i q^i\right]
 \, \cdot \,  \exp \left[ {\rho D} \right]
\label{simplecostrepresentative}
\end{eqnarray}
This amounts to the following coordinate change:
\begin{eqnarray}
z&=& \left( 1-\ft12 \rho + \ft16 \rho^2 + {\cal O}(\rho^3) \right)\,
x \,, \nonumber \\
\xi^i&=& \left(1-\ft14 \rho + \ft{1}{24} \rho^2 + {\cal O}(\rho^3)\right)\,
\theta^i \,.
\label{changecos}
\end{eqnarray}
This is the parametrization that was used in
\cite{torinos7} to get the $Osp(8\vert 4)$-singleton action
from the supermembrane. For this choice of coordinates the anti de Sitter metric
takes the standard form (\ref{adsmet}).
The Killing vectors are
\begin{eqnarray}
\veck[ P_m ]&=& -i\, \partial_m \,, \nonumber \\
\veck [q^{\alpha i}] &=&
\frac{\partial}{\partial \xi_\alpha^i}
-\frac{1}{2}\left(\gamma^m\xi^i\right)^\alpha
\partial_m \,,
\nonumber \\
\veck [J^m] &=& \varepsilon^{mpq} z_p \partial_q
-\frac{i}{2} \left(\xi^i\gamma^m \right)_\alpha \frac{\partial}{\partial
\xi_\alpha^i}
\,, \nonumber \\
\veck [D] &=& \frac{\partial}{\partial \rho}
- z \cdot \partial
-\frac{1}{2}\xi_\alpha^i
\frac{\partial}{\partial \xi_\alpha^i}
\,,
\nonumber \\
\veck [s^{\alpha i}] &=&
-\xi^{\alpha i} \frac{\partial}{\partial \rho}
+\frac{1}{2}  \xi^{\alpha i}\, z \cdot \partial
+ \frac{i}{2} \varepsilon^{pqm}\, z_p(\gamma_q \xi^i)^\alpha \partial_m
\nonumber \\
& &
-\frac{1}{8} (\xi^j \xi^j) (\gamma^m \xi^i)^\alpha \partial_m
- z^m (\gamma_m)^\alpha{}_\beta \frac{\partial}{\partial \xi^i_\beta}
-\frac{1}{4} (\xi^j\xi^j) \frac{\partial}{\partial \xi^i_\alpha}
\nonumber \\
& &
+\frac{1}{2} \xi^{\alpha i} \xi^{\beta j}
\frac{\partial}{\partial\xi^j_\beta}
-\frac{1}{2}(\gamma^m \xi^i)^\alpha \xi^j_\beta \gamma_m
\frac{\partial}{\partial\xi^j_\beta}
\,.
\label{simplekilling}
\end{eqnarray}
and for the compensators we find:
\begin{eqnarray}
W[P]&=& 0 \,, \nonumber \\
W[q^{\alpha i}] &=& 0 \,, \nonumber \\
W[J^m] &=& J^m \,, \nonumber \\
W[D] &=& 0 \,, \nonumber \\
W[s^{\alpha i}] &=&  s^{\alpha i}
  - i\, \left(\gamma^m \theta^i \right)^\alpha \,J_m
  + i  \theta^{\alpha j} \, T^{ij} \,.
\label{compensatorsAdSrzxi}
\end{eqnarray}
For a detailed
derivation of these Killing vectors and compensators we refer
the reader to  appendix \ref{derivationkillings}.
\par
The boundary superspace $\partial(AdS_{4\vert 2{\cal N}})$ is formed by the
points on the
supercoset with $\rho=0$:
\begin{eqnarray}
L^{CFT}(y) = \exp \left[\mbox{ i}\, x \cdot P + \theta^i q^i\right]
\label{boundarycosetrepresentative}
\end{eqnarray}
In order to see how the supergroup acts on fields that
live on this boundary we  use the fact that
this submanifold is by itself a supercoset. Indeed
instead of $H^{AdS} \subset Osp({\cal N} \vert 4)$ as given in  (\ref{HAdS}), we
can choose the larger subalgebra
\begin{equation}
H^{CFT} = \mbox{span} \, \left \{ D, J^m, K_m, s_\alpha^{i}, T^{ij} \right \}
\,.
\label{HCFT}
\end{equation}
and consider the new supercoset $G/H^{CFT}$.
By defintion also on this smaller space we have a non linear realization of the
full orthosymplectic superalgebra.
For the Killing vectors we find:
\begin{eqnarray}
\veck [P_m] &=& -i\, \partial_m \,, \nonumber \\
\veck [q^{\alpha i}] &=&
\frac{\partial}{\partial \theta_\alpha^i}
-\frac{1}{2}\left(\gamma^m\theta^i\right)^\alpha
\partial_m \,,
\nonumber \\
\veck [J^m] &=& \varepsilon^{mpq} x_p \partial_q
-\frac{i}{2} \left(\theta^i\gamma^m \right)_\alpha \frac{\partial}{\partial
\theta_\alpha^i}
\,, \nonumber \\
\veck[D] &=&
- x \cdot \partial
-\frac{1}{2}\theta_\alpha^i
\frac{\partial}{\partial \theta_\alpha^i} \,,
\nonumber \\
\veck [s^{\alpha i}] &=&
\frac{1}{2} \theta^{\alpha i}\, x \cdot \partial
+ \frac{i}{2} \varepsilon^{pqm}\, x_p(\gamma_q \theta^i)^\alpha \partial_m
-\frac{1}{8} (\theta^j \theta^j) (\gamma^m \theta^i)^\alpha \partial_m
\nonumber \\
& &
- x^m (\gamma_m)^\alpha{}_\beta \frac{\partial}{\partial \theta^i_\beta}
-\frac{1}{4} (\theta^j\theta^j) \frac{\partial}{\partial \theta^i_\alpha}
+\frac{1}{2} \theta^{\alpha i} \theta^{\beta j}
\frac{\partial}{\partial\theta^j_\beta}
-\frac{1}{2}(\gamma^m \theta^i)^\alpha \theta^j_\beta \gamma_m
\frac{\partial}{\partial\theta^j_\beta}
\,.  \nonumber\\
\label{bounkil}
\end{eqnarray}
and for the compensators we have:
\begin{eqnarray}
W[P_m] &=& 0 \,, \nonumber \\
W[q^{\alpha i}] &=& 0 \,, \nonumber \\
W[J^m] &=& J^m \,, \nonumber \\
W[D] &=& D \,, \nonumber\\
W[s^{\alpha i}] &=& - \theta^{\alpha i}\, D +
     s^{\alpha i} - i\, \left( \gamma^m \theta^i \right)^\alpha\,J_m
     + i \theta^j T^{ij} \,.
\label{compensatorsdAdS}
\end{eqnarray}
If we compare the Killing vectors on the boundary (\ref{bounkil})  with those on
the
bulk (\ref{simplekilling}) we see that they are very similar. The
only formal difference is the suppression of the $\frac{\partial}{\partial\rho
}$ terms. The conceptual difference, however, is relevant. On the
boundary the transformations generated by (\ref{bounkil}) are the
{\it standard superconformal transformations} in three--dimensional
(compactified) Minkowski space.
In the bulk the transformations generated by (\ref{simplekilling})
are {\it superisometries} of anti de Sitter superspace. They might be
written in completely different but equivalent forms if we used other
coordinate frames. The form they have is due to the use of the {\it
solvable coordinate frame} $(\rho, z, \xi)$ which is the most appropriate to study the
restriction of bulk supermultiplets to the boundary.
For more details on this point we refer the reader to appendix
\ref{derivationkillings}
%
%
\section{$Osp(2\vert 4)$ superfields in the bulk and on the boundary}
\label{supfieldbuetbo}
As we explained in the introduction our main goal is the determination
of the ${\cal N}=2$ three dimensional gauge theories associated with the
sasakian horizons (\ref{sasaki}) and the comparison between
Kaluza Klein spectra of M--theory compactified on $AdS_4$ times such
horizons with the spectrum of primary conformal superfields of
the corresponding gauge theory. For this reason
we mainly focus on the case of $Osp(2\vert 4)$ supermultiplets.
As already stressed  the structure of such supermultiplets
has been determined in Kaluza Klein language in \cite{multanna, m111spectrum}.
Hence they have been obtained in the basis (\ref{ospE}) of the
orthosymplectic superalgebra.
Here we consider their translation into the superconformal language
provided by the other basis (\ref{ospD}). In this way we will
construct a boundary superfield associated with each particle
supermultiplet of the bulk. The components of the supermultiplet
are Kaluza Klein states: it follows that we obtain a one--to--one
correspondence between Kaluza Klein states and components of the
boundary superfield.
\subsection{Conformal $Osp(2 \vert 4)$ superfields: general discussion}
So let us restrict our attention to ${\cal N}\!\!=\!\!2$.
In this case the $SO(2)$ group has just one generator that we name
the hypercharge:
\begin{equation}
Y \equiv T^{21} \,.
\end{equation}
Since it is convenient to work with eigenstates of the
hypercharge operator, we reorganize the two Grassman spinor coordinates of
superspace in complex combinations:
\begin{eqnarray}
\theta^\pm_\alpha = \frac{1}{\sqrt{2}}(\theta^1_\alpha
\pm i \theta^2_\alpha) \,,
\qquad Y \, \theta^\pm_\alpha = \pm \theta^\pm_\alpha
\label{complexthet}
\end{eqnarray}
In this new notations the Killing vectors generating
$q$--supersymmetries on the boundary (see eq.(\ref{bounkil})) take the
form:
\begin{equation}
{\vec k}\left[q^{\alpha i}\right]  \quad \longrightarrow \quad q^{\alpha \pm}=
\frac{\partial}{\partial \theta^\mp_\alpha}
- \ft{1}{2} \,  (\gamma^m)^\alpha{}_\beta \theta^{\beta \pm} \partial_m \,,
\label{qpm}
\end{equation}
A generic superfield is a function $\Phi(x,\theta)$ of the bosonic
coordinates $x$ and of all the $\theta .s$ Expanding such a field in
power series of the $\theta .s$ we obtain a multiplet of $x$--space
fields that, under the action of the Killing vector (\ref{qpm}), form a
representation of Poincar\'e supersymmetry. Such a representation can
be shortened by imposing on the superfield $\Phi(x,\theta)$
constraints that are invariant with respect to the action of the
Killing vectors (\ref{qpm}). This is possible because of the
existence of the so called superderivatives, namely of fermionic vector
fields that commute with the supersymmetry Killing vectors. In our
notations the superderivatives are defined as follows:
\begin{eqnarray}
{\cal D}^{\alpha \pm} =
\frac{\partial}{\partial \theta^\mp_\alpha}
+\ft{1}{2} \,  (\gamma^m)^\alpha{}_\beta \theta^{\beta \pm} \partial_m \,,
\label{supdervcf}
\end{eqnarray}
and satisfy the required property
\begin{equation}
\begin{array}{rclcr}
\{{\cal D}^{\alpha \pm}, q^{\beta \pm}\}&=&
\{{\cal D}^{\alpha \pm}, q^{\beta \mp}\}&=&0 \,.
\end{array}
\label{commDq}
\end{equation}
As explained in \cite{castdauriafre} the existence of
superderivatives is the manifestation at the fermionic level of a
general property of coset manifolds. For $G/H$ the true isometry
algebra is not $G$, rather it is $G \times N(H)_G$ where $N(H)_G$
denotes the normalizer of the stability subalgebra $H$. The
additional isometries are generated by {\it right--invariant} rather
than {\it left--invariant} vector fields that as such commute with
the {\it left--invariant} ones. If we agree that the Killing vectors
are left--invariant vector fields than the superderivatives are
right--invariant ones and generate the additional superisometries of
Poincar\'e superspace. Shortened representations of Poincar\'e
supersymmetry are superfields with a prescribed behaviour under the
additional superisometries: for instance they may be invariant under
such transformations. We can formulate these shortening conditions by
writing constraints such as
\begin{equation}
  {\cal D}^{\alpha +}\Phi(x,\theta)=0 \, .
\label{typconstr}
\end{equation}
The key point in our  discussion is that a constraint of type
(\ref{typconstr}) is guaranteed from eq.s (\ref{commDq}) to be
invariant with respect to the superPoincar\'e algebra, yet it is not
a priori guaranteed that it is invariant under the action of the full
superconformal algebra (\ref{bounkil}). Investigating the additional
conditions that make a constraint such as (\ref{typconstr})
superconformal invariant is the main goal of the present section.
This is the main tool that allows a transcription of the
Kaluza--Klein results for supermultiplets into a superconformal
language.
\par
To develop such a programme it is useful to perform a further
coordinate change that is quite traditional in superspace literature.
Given the coordinates  $x$ on the boundary (or the coordinates $z$ for
the bulk) we set:
\begin{eqnarray}
y^m=x^m + \ft{1}{2} \, \theta^+ \gamma^m \theta^- \,.
\end{eqnarray}
Then the superderivatives become
\begin{eqnarray}
{\cal D}^{\alpha +} &=&  \frac{\partial}{\partial \theta^-_\alpha}
\,, \nonumber \\
{\cal D}^{\alpha -} &=& \frac{\partial}{\partial \theta^+_\alpha}
+ (\gamma^m)^\alpha{}_\beta\theta^{\beta -}
\partial_m \,.
\end{eqnarray}
It is our aim to describe superfield multiplets both on the bulk
and on the boundary. It is clear that one can do the
same redefinitions for the Killing vector of $q$-supersymmetry (\ref{qpm})
and that  one can introduce superderivatives also for the theory on the
bulk. In that case one inserts the functions $t(\rho)$ and
$\g(\rho)$ in the above formulas or if one uses the solvable coordinates
$(\rho, z, \xi)$ as in (\ref{changecos}) then there is just no difference
with the boundary case.
\par
So let us finally turn to superfields.
 We begin by  focusing on boundary
superfields since their treatment is
slightly easier than  the treatment of bulk superfields.
\bdefi
A primary superfield is defined as follows
(see \cite{gunaydinminiczagerman2, macksalam}),
\begin{eqnarray}
\Phi^{\partial AdS}(x, \theta) =\exp \left [\mbox{i}\, x\cdot P + \theta^i
q^i\right ]
 \Phi(0)\,,
\label{3Dsuperfield}
\end{eqnarray}
where $\Phi(0)$ is a primary state (see eq.(\ref{primstate}))
\begin{eqnarray}
s_\alpha^i \Phi(0) &=& 0 \,, \nonumber \\
K_m \Phi(0) &=& 0 \,.
\end{eqnarray}
of  scaling weight $D_0$, hypercharge
$y_0$ and eigenvalue $j$ for the ``third-component'' operator $J_2$
\begin{equation}
\begin{array}{ccccc}
  D \, \Phi(0) = D_0 \, \Phi(0)&;& Y \, \Phi(0) = y_0 \, \Phi(0) &;&
J_2\, \Phi(0) = j \, \Phi(0)
\end{array}
\label{primlab}
\end{equation}
\edefi
From the above defintion one sees that the primary superfield
$\Phi^{\partial AdS}(x, \theta)$ is
actually obtained by acting with the coset representative
(\ref{boundarycosetrepresentative}) on the
$SO(1,2)\times SO(1,1)$-primary state.
Hence we know how it transforms under the infinitesimal
transformations of the group $Osp(2\vert 4)$. Indeed one simply uses
(\ref{infinitesimalcosettransfo}) to obtain the result.
For example under dilatation we have:
\begin{eqnarray}
D\, \Phi^{\partial AdS}(x,\theta) = \left(-x\cdot\partial - \ft12 \theta^i
\frac{\partial}{\partial \theta^i} + D_0   \right) \Phi(x,\theta),
\end{eqnarray}
where the term $D_0$ comes from the compensator in
(\ref{compensatorsdAdS}).
Of particular interest is the transformation under
special supersymmetry since it imposes the constraints
for shortening,
\begin{equation}
s^\pm \Phi^{\partial AdS}(x,\theta) = \veck [s^\pm] \Phi(x, \theta)
                      + e^{i\, x\cdot P + \theta^i q^i}
                       \left( -\theta^\pm D
                              - i\, \gamma^m \theta^\pm \,J_m
                              +s^\pm
                              \pm \theta^\pm Y \right) \Phi(0)
                      \,. \label{spm}
\end{equation}
For completeness we give the form
of $s^\pm$ in the $y$-basis where it gets a relatively concise form,
\begin{eqnarray}
\veck [s^{\alpha-}] &=& - \left( y\cdot \gamma\right)^\alpha{}_\beta
\frac{\partial}{\partial \theta_\beta^+}
+ \frac{1}{2} \left( \theta^- \theta^- \right)
\frac{\partial}{\partial \theta^-_\alpha}
\nonumber \\
\veck [s^{\alpha+}] &=& \theta^{\alpha+} y\cdot\partial
+ i\, \varepsilon^{pqm} y_p \left( \gamma_p \theta^+ \right)^\alpha\partial_m
+ \frac{1}{2} \left(\theta^+ \theta^+\right)
\frac{\partial}{\partial \theta^+_\alpha}
\nonumber \\
& & + \theta^+\gamma^m \theta^- \left(\gamma_m\right)^\alpha{}_\beta
\frac{\partial}{\partial \theta_\beta^-}
\,,
\label{finalkillings}
\end{eqnarray}
\par
Let us now turn to a direct discussion of multiplet shortening and
consider the superconformal invariance of Poincar\'e constraints
constructed with the superderivatives ${\cal D}^{\alpha \pm}$. The
simplest example is provided by the {\it chiral supermultiplet}.
By definition this is a scalar superfield $\Phi_{chiral}(y,\theta)$
obeying the constraint (\ref{typconstr}) which is solved by boosting
only along $q^-$ and not along $q^+$:
\begin{eqnarray}
\Phi_{chiral}(y,\theta)=e^{i\, y \cdot P + \theta^+ q^-} \Phi(0),
\end{eqnarray}
Hence we have
\begin{eqnarray}
\Phi_{chiral}(\rho, y, \theta) = X(\rho, y) + \theta^+ \lambda(\rho, y) +
\theta^+
\theta^+ H(\rho, y)
\end{eqnarray}
on the bulk or
\begin{eqnarray}
\Phi_{chiral}(y, \theta) = X(y) + \theta^+ \lambda(y) + \theta^+
\theta^+ H(y)
\label{chiralmultiplet3D}
\end{eqnarray}
on the boundary. The field components of the chiral multiplet are:
\begin{eqnarray}
X = e^{i\, y\cdot P} \, \Phi(0)\,,\qquad
\lambda = i \, e^{i\, y\cdot P} \,  q^- \Phi(0) \,, \qquad
H = -\ft14 e^{i\, y\cdot P} \, q^- q^- \phi(0)
\,.
\label{chiralcomponents}
\end{eqnarray}
For completeness,
we write the superfield $\Phi$ also in the
$x$-basis\footnote{where $\Box=\partial^m \partial_m\,$.},
\begin{eqnarray}
\Phi(x) &=& X(x)
+\theta^+ \lambda(x) + (\theta^+\theta^+) H(x)
+\ft{1}{2} \theta^+ \gamma^m \theta^- \partial_m X(x)
\nonumber \\
& &
+\ft{1}{4}  (\theta^+\theta^+) \theta^- \dslash \lambda(x)
+\ft{1}{16} (\theta^+ \theta^+) (\theta^- \theta^-)
\Box X(x)
\nonumber \\
&=& \exp\left(\ft{1}{2} \theta^+ \gamma^m\theta^-
\partial_m \right) \Phi(y) \,.
\label{formchiral}
\end{eqnarray}
\par
Because of (\ref{commDq}), we are guaranteed that under $q$--supersymmetry
the chiral superfield
$\Phi_{chiral}$ transforms into a chiral superfield. We should  verify
that this is true also for $s$--supersymmetry.
To say it simply we just have to check that   $s^- \Phi_{chiral}$ does not
depend
on $\theta^-$. This is not generically true, but it becomes true
if certain  extra conditions on the quantum numbers of the primary
state are satisfied. Such conditions are the same one obtains as
multiplet shortening conditions when constructing the UIR.s of the
superalgebra with the {\it norm method} of Freedman and Nicolai
\cite{freedmannicolai}
or with the {\it oscillator method} of G\"unaydin and collaborators
\cite{gunay2,gunawar,gunaydinminiczagerman1, gunaydinminiczagerman2}
\footnote{We are particularly grateful to S. Ferrara for explaining
to us this general idea that, extended from the case of $AdS_5/CFT_4$
to the case $AdS_4/CFT_3$, has been an essential guiding line in the
development of the present work}.
\par
In the specific instance of the chiral multiplet,
looking at (\ref{spm}) and (\ref{finalkillings})
we see that  in $s^-\Phi_{chiral}$ the terms  depending on $\theta^-$
are the following ones:
\begin{eqnarray}
s^- \Phi \Big\vert_{\theta^-} = - \left( D_0+ y_0 \right)  \theta^- \Phi =0 \,,
\end{eqnarray}
they cancel if
\begin{equation}
D_0 = - y_0 \,.
\label{Disy}
\end{equation}
Eq.(\ref{Disy}) is easily recognized as the unitarity condition for
the existence of $Osp(2\vert 4)$ hypermultiplets (see
\cite{multanna,m111spectrum}).
The algebra (\ref{Kiss2}) ensures that the chiral multiplet
also transforms into a chiral multiplet under $K_m$. Moreover
we know that the action of the compensators of $K_m$ on
the chiral multiplet is zero. Furthermore,
the compensators of the generators $P_m, q^i, J_m$
on the chiral multiplet are zero and from (\ref{infinitesimalcosettransfo})
we conclude that their generators act on the chiral multiplet
as the Killing vectors.
\par
Notice that the linear part of the
$s$-supersymmetry transformation on the chiral multiplet
has the same form of the $q$-supersymmetry but with the
parameter taken to be $\epsilon_q = -i\, y \cdot \g \epsilon_s$.
As already stated the non-linear form of  $s$-supersymmetry
is the consequence of its gauge fixing which we have implicitly imposed
from the start by choosing the supersolvable Lie algebra
parametrization of superspace and by  taking the coset representatives
as in (\ref{AdScosetrepresentative}) and (\ref{boundarycosetrepresentative}).
\footnote
{  Just as a comment
we recall that the standard way of gauge fixing  special supersymmetry
in a superconformal theory is to impose a gauge-fixing
condition and then modify  $q$-supersymmetry
by means of a decomposition rule, i.e. adding to it
special supersymmetry with specific parameters that depend
on the supersymmetry parameters, such that the gauge-fixing
condition becomes invariant under the modified supersymmetry.
In our case we still have the standard form of $q$-supersymmetry
but upon gauge fixing $s$-supersymmetry has become
non-linear. The fact that  $s$-supersymmetry partly
resembles  $q$-supersymmetry comes from the fact that
it can be seen as a $q$-like supersymmetry
with its own superspace coordinates, which upon gauge fixing
have become dependent on the $\theta$-coordinates.}
In addition to the chiral multiplet there exists also
the complex conjugate {\it antichiral multiplet}
${\bar \Phi}_{chiral}=\Phi_{antichiral}$
with opposite hypercharge and the relation $D_0=y_0$.
\subsection{Matching the Kaluza Klein results for $Osp(2 \vert 4)$
supermultiplets with boundary conformal superfields}
It is now our purpose to reformulate the ${\cal N}=2$ multiplets
found in Kaluza Klein supergravity \cite{m111spectrum} in terms of
superfields living on the boundary of the $AdS_4$ space--time manifold.
This is the key step to convert information coming from classical
harmonic analysis on the compact manifold $X^7$ into predictions on
the spectrum of conformal primary operators present in the
three--dimensional gauge theory of the M2--brane.
Although the results obtained in \cite{m111spectrum} refer to a
specific case, the structure of the multiplets is general and applies
to all ${\cal N}=2$ compactifications, namely to all sasakian
horizons $X^7$. Similarly general are the recipes discussed in the
present section to convert Kaluza--Klein data into boundary
superfields.
\par
As shown in \cite{m111spectrum} there are three types of {\bf long
multiplets} with the following {\it bulk spin content}:
\begin{enumerate}
  \item The long graviton multiplet $~~\left(1\left(2\right),
  4\left(3\over 2\right),6\left(1\right),4\left(1\over 2\right),
  1\left(0\right)\right)$
  \item The long gravitino multiplet $~~\left(1\left({3\over 2}\right),4\left(1\right),
  6\left(1\over 2\right),4\left(0\right)\right)$
  \item The long vector multiplets
  $~~\left(1\left(1\right),4\left(1\over 2\right),5\left(0\right)\right)$
\end{enumerate}
and four types of {\bf short multiplets} with the following {\it bulk spin content}:
\begin{enumerate}
  \item the short graviton multiplet
  $~~\left(1\left(2\right),3\left(3\over 2\right),3\left(1\right),
  1\left({1\over 2}\right)\right)$
  \item the short gravitino multiplet
  $~~\left(1\left({3\over 2}\right),3\left(1\right),3\left(1\over 2\right),
  1\left(0\right)\right)$
  \item the short vector multiplet
  $~~\left(1\left(1\right),3\left(1\over 2\right),3\left(0\right)\right)$
  \item the hypermultiplet  $~~\left(2\left(1\over 2\right),4\left(0\right)\right)$
\end{enumerate}
Finally there are the {\bf ultrashort multiplets} corresponding to
the massless multiplets available in ${\cal N}=2$ supergravity
and having the following {\it bulk spin content}:
\begin{enumerate}
  \item the massless graviton multiplet $~~\left(1\left(2\right),
  2\left({3\over 2}\right),1\left(1\right)\right)$
  \item the massless vector multiplet
   $~~\left(1\left(1\right),2\left(\ft{1}{2}\right),2\left(0\right)\right)$
\end{enumerate}
Interpreted as superfields on the boundary the  {\it long multiplets}
correspond to {\it unconstrained superfields} and their discussion is quite
straightforward.
We are mostly interested in short multiplets that correspond to
composite operators of the microscopic gauge theory with protected
scaling dimensions. In superfield language, as we have shown in the
previous section, {\it short multiplets} are constrained superfields.
\par
Just as on the boundary, also in the bulk, we obtain such constraints
by means of the bulk superderivatives. In order to show how
this works we begin  by discussing the {\it chiral superfield  in the
bulk} and then show how it is obtained from the hypermultiplet
found in Kaluza Klein theory \cite{m111spectrum}.
\subsubsection{Chiral superfields are the Hypermultiplets: the basic example}
The treatment for the bulk chiral field is completely analogous
to that of chiral superfield on the boundary.
\par
Generically bulk superfields are given by:
\begin{eqnarray}
\Phi^{AdS}(\rho, x, \theta) = \exp \left[\rho D + \mbox{i}\, x\cdot P +
 \theta^i q^i\right ]
\, \Phi(0)\,,
\label{4Dsuperfield}
\end{eqnarray}
Using the parametrization (\ref{changecos}) we can rewrite
(\ref{4Dsuperfield}) in the following way:
\begin{eqnarray}
\Phi^{AdS}(\rho, z, \xi) = \exp \left[\mbox{i}\, z \cdot P + \xi^i q^i\right ]
                 \, \cdot \, \exp \left[\rho D_0\right ]\, \Phi(0)\,.
\label{simpleAdSsuperfield}
\end{eqnarray}
Then the  generator $D$ acts on this field as follows:
\begin{eqnarray}
D \, \Phi^{AdS}(\rho, z, \xi) =
\left(- z \cdot \partial
      - \ft12
        \xi^i \frac{\partial}{\partial \xi^i}
        + D_0
\right) \Phi^{AdS}(\rho, z, \xi) \,.
\end{eqnarray}
Just as for boundary  chiral superfields, also in the bulk  we find that
the constraint (\ref{typconstr}) is
invariant under   the $s$-supersymmetry rule (\ref{simplekilling}) if
and only if:
\begin{equation}
  D_0= - y_0
\label{dugaly}
\end{equation}
Furthermore, looking at (\ref{simpleAdSsuperfield})
one sees that for the bulk superfields $D_0=0$ is forbidden.
This constraint on the scaling dimension
together with the relation $E_0=-D_0$, coincides with the
constraint:
\begin{equation}
  E_0=\vert y_0 \vert
\label{Eugaly}
\end{equation}
defining the $Osp(2\vert 4)$ hypermultiplet UIR of $Osp(2\vert 4)$
constructed with the norm method and in the formulation (\ref{ospE})
of the superalgebra (see \cite{multanna, m111spectrum}).
The transformation of the bulk chiral superfield under
$s, P_m, q^i, J_m$ is simply given by the bulk
Killing vectors. In particular the form of the $s$-supersymmetry Killing vector
coincides with that given in
(\ref{finalkillings}) for the boundary.
\par
As we saw a chiral superfield in the bulk
describes an $Osp(2\vert 4)$
hypermultiplet. To see this explicitly it suffices to look at the following
table\footnote{The hypercharge $y_0$ in the table is chosen to
be positive.}
\begin{eqnarray}
\centering
\begin{array}{||c|c|c||}
\hline
{\rm Spin} &
{\rm Particle \, \, states}     &
{\rm Name}  \\
\hline
\hline
 & &  \\
\ft12  & \bar a^- \vert E_0=y_0 , y_0 \rangle  & \lambda_L   \\
0      & \bar a^- \bar a^- \vert E_0=y_0, y_0 \rangle  & \pi     \\
0      & \vert E_0=y_0, y_0 \rangle  & S       \\
 & &  \\
\hline
\hline
 & &  \\
\ft12  & \bar a^+ \vert E_0=y_0, -y_0 \rangle & \lambda_L   \\
0      & \bar a^+ \bar a^+ \vert E_0=y_0, -y_0 \rangle & \pi     \\
0      & \vert E_0=y_0, -y_0 \rangle & S       \\
 & &  \\
\hline
\end{array}
\label{hyper}
\end{eqnarray}
where we have collected  the particle states  forming
a hypermultiplet as it appears in Kaluza Klein supergravity
on $AdS_4 \times X^7$, whenever $X^7$ is sasakian.
The names of the fields are the standard ones introduced in
\cite{univer} for the
linearization of D=11 supergravity on $AdS_4 \times X^7$ and used
in \cite{multanna, m111spectrum}.
Applying the rotation matrix $U$  of eq. (\ref{rotationmatrix})
to the states in the upper part of this table
 we  indeed find the field components (\ref{chiralcomponents}) of
the chiral supermultiplet.
\par
Having clarified how to obtain the four-dimensional
chiral superfield from the $Osp(2\vert 4)$ hypermultiplet
we can now obtain the other shortened $Osp(2\vert 4)$
superfields from the information that was obtained in
\cite{m111spectrum}.
In \cite{m111spectrum} all the field components of the
$Osp(2\vert 4)$ multiplets were listed together with
their spins $s_0$, energy $E_0$ and their   hypercharge
$y_0$. This is sufficient to reconstruct the particle
states of the multiplets which are given by the states
(\ref{energyreps}). Indeed the energy determines the
number of energy boosts that are applied to the vacuum
in order to get the state. The hypercharge determines
the number of $\bar a^+$ and/or $\bar a^-$ present. Finally
$s_0$ tells us what spin we should get. In practice this means
that we always have to take the symmetrization of the
spinor indices, since $(\alpha_1\dots\alpha_n)$ yields
a spin-$\ft{n}{2}$ representation. Following \cite{multanna}
we ignore the unitary representations of $SO(3,2)$ that
are built by the energy boosts $M_i^+$ and we just list the ground
states for each UIR of $SO(2,3)$ into which the UIR of $Osp(2\vert
4)$ decomposes.
\subsubsection{Superfield description of the short vector multiplet}
Let us start with the short massive vector multiplet. From
\cite{m111spectrum} we know that the constraint for shortening is
\begin{equation}
  E_0=\vert y_0 \vert + 1
\label{eisyplusone}
\end{equation}
and that the particle states
of the multiplet are given by,\footnote{The
hypercharge $y_0$ in the table is chosen to
be positive.}
\begin{eqnarray}
\label{shvecmu}
\centering
\begin{array}{||c|c|c||}
\hline
{\rm  Spin} &
{\rm Particle \, \, states}     &
{\rm Name}  \\
\hline
\hline
 & &  \\
1      & ( \bar a^- \tau^i \bar a^+ ) \vert E_0=y_0+1,  y_0 \rangle  & A   \\
\ft12  & ( \bar a^- \bar a^- ) \bar a^+ \vert y_0+1,  y_0 \rangle  & \lambda_T \\
\ft12  & \bar a^+ \vert y_0+1,  y_0 \rangle  & \lambda_L  \\
\ft12  & \bar a^- \vert y_0+1,  y_0 \rangle  & \lambda_L  \\
0      & ( \bar a^- \bar a^- ) \vert y_0+1,  y_0 \rangle  & \pi \\
0      & ( \bar a^+ \bar a^- ) \vert y_0+1,  y_0 \rangle  & \pi \\
0      &    \vert y_0+1,  y_0 \rangle        &  S  \\
 & &  \\
\hline
\hline
 & &  \\
1  & (\bar a^+ \tau^i \bar a^-) \vert E_0=y_0+1, - y_0 \rangle  & A   \\
\ft12  & ( \bar a^+ \bar a^+ ) \bar a^- \vert y_0+1, - y_0 \rangle  & \lambda_T \\
\ft12  & \bar a^- \vert y_0+1, - y_0 \rangle  & \lambda_L  \\
\ft12  & \bar a^+ \vert y_0+1, - y_0 \rangle  & \lambda_L  \\
0      & ( \bar a^+ \bar a^+ ) \vert y_0+1, - y_0 \rangle  & \pi \\
0      & ( \bar a^+ \bar a^- ) \vert y_0+1, - y_0 \rangle  & \pi \\
0      &    \vert y_0+1, - y_0 \rangle        &  S  \\
 & &  \\
\hline
\end{array}
\end{eqnarray}
where we have multiplied the symmetrized product $\bar a^+_\alpha \bar a^-_\beta$
 with the $\tau$-matrices in order
to single out the $SO(3)$ vector index $i$ that labels the on--shell states of the
$d=4$ massive vector field.
Applying the rotation matrix $U$ to the states in the upper part of
 table (\ref{shvecmu}) we find the following states:
\begin{eqnarray}
S=\vert {\rm vac} \rangle \,, \qquad
\lambda^{\pm}_L = i \, q^\pm \vert {\rm vac} \rangle \,, \qquad
\pi^{--} = -\ft14 \, q^- q^- \vert {\rm vac} \rangle \,, \qquad
{\rm etc} \dots
\label{rotatedstates}
\end{eqnarray}
where we used the same notation for the rotated
as for the original states
and up to an irrelevant factor $\ft14$.
We  follow the same procedure   also for the other short and
massless multiplets. Namely in the superfield transcription
of our multiplets we  use the same
names for the superspace field components as
 for the particle fields appearing  in the
$SO(3)\times SO(2)$ basis. Moreover when convenient we rescale some field
components without mentioning it explicitly.
The list of states appearing in  (\ref{rotatedstates})
are the components of a superfield
\begin{eqnarray}
\Phi_{vector} &=& S + \theta^- \lambda_L^+ + \theta^+ \lambda_L^-
                + \theta^+ \theta^- \pi^0
                + \theta^+ \theta^+ \pi^{--}
                + \theta^+ \Aslash \theta^-
                + \theta^+ \theta^+ \, \theta^- \lambda_T^- \,,
\nonumber \\ & &
\label{vecsupfil}
\end{eqnarray}
which is the explicit solution of the following constraint
\begin{equation}
{\cal D}^+ {\cal D}^+ \Phi_{vector} = 0 \,.
\end{equation}
imposed on a superfield of the form (\ref{4Dsuperfield})
with hypercharge $y_0$.
\par
In superspace literature a superfield of type
(\ref{vecsupfil}) is named a linear superfield.
If we consider the variation of a linear superfield with respect to
 $s^-$, such variation contains, a priori, a term of the form
\begin{equation}
s^- \Phi_{vector} \Big\vert_{\theta^-\theta^-}=
\ft12 \left(D_0 + y_0+1\right)(\theta^-\theta^-) \lambda_L^+ \,,
\end{equation}
which has to cancel if $\Phi_{vector}$ is to transform into
a linear multiplet  under $s^-$. Hence the following condition has to
be imposed
\begin{equation}
D_0 = - y_0 - 1 \,.
\label{dismymone}
\end{equation}
which is identical with the bound for the vector multiplet
shortening $E_0=y_0+1$ found in \cite{multanna,m111spectrum}.
\subsubsection{Superfield description of the short gravitino
multiplet}
Let us consider the short gravitino multiplets found in \cite{m111spectrum}.
The particle state content of these multiplet is given below\footnote{The
hypercharge $y_0$ in the table is chosen to
be positive.}:
\begin{eqnarray}
\centering
\begin{array}{||c|c|c||}
\hline
{\rm  Spin} &
{\rm Particle \, \, states}     &
{\rm Name}  \\
\hline
\hline
 & &  \\
\ft32      &  \bar a^-_{(\alpha} \bar a^+_{\beta}
\vert E_0=y_0+\ft32,  s_0=\ft12, y_0 \rangle_{\gamma)}  & \chi^{(+)} \\
1  & (\bar a^- \bar a^-) \bar a^+\tau^i \vert y_0+\ft32,  \ft12, y_0 \rangle & Z \\
1  & \bar a^+ \tau^i \vert y_0+\ft32,  \ft12, y_0 \rangle  & A  \\
1  & \bar a^-  \tau^i \vert y_0+\ft32,  \ft12, y_0 \rangle & A  \\
\ft12      & ( \bar a^- \bar a^- ) \vert y_0+\ft32,  \ft12, y_0 \rangle  & \lambda_T \\
\ft12      &  (3 \,\bar a^- \bar a^+ + \bar a^- \tau^i \bar a^+\,  \tau_i )
 \vert y_0+\ft32,  \ft12, y_0 \rangle  & \lambda_T \\
\ft12      &    \vert y_0+\ft32,  \ft12, y_0 \rangle       &  \lambda_L  \\
0  & \bar a^- \vert y_0+\ft32,  \ft12, y_0 \rangle  & \phi \\
 & &  \\
\hline
\hline
 & &  \\
\ft32      &  \bar a^+_{(\alpha} \bar a^-_{\beta}
\vert E_0=y_0+\ft32, s_0=\ft12, -y_0 \rangle_{\gamma)},  & \chi^{(+)} \\
1  & (\bar a^+ \bar a^+) \bar a^-\tau^i \vert
y_0+\ft32, \ft12, -y_0 \rangle & Z \\
1  & \bar a^- \tau^i \vert y_0+\ft32, \ft12, -y_0 \rangle  & A  \\
1  & \bar a^+  \tau^i \vert y_0+\ft32, \ft12, -y_0 \rangle & A  \\
\ft12      & ( \bar a^+ \bar a^+ ) \vert y_0+\ft32, \ft12, -y_0 \rangle  & \lambda_T \\
\ft12      &  (3 \,\bar a^+ \bar a^- + \bar a^+ \tau^i \bar a^-\,  \tau_i )
 \vert y_0+\ft32, \ft12, -y_0 \rangle  & \lambda_T \\
\ft12      &    \vert y_0+\ft32, \ft12, -y_0 \rangle       &  \lambda_L  \\
0  &  \bar a^+ \vert y_0+\ft32, \ft12, -y_0 \rangle  & \phi \\
 & &  \\
\hline
\end{array}
\label{shgrinomu}
\end{eqnarray}
Applying the rotation matrix $U$ (\ref{rotationmatrix})
to the upper part of table (\ref{shgrinomu}),
and identifying the particle states with the corresponding rotated
field states as we have done in the previous cases,
we find  the following spinorial
superfield
\begin{eqnarray}
\Phi_{gravitino} &=& \lambda_L + \Aplusslash \theta^- + \Aminusslash \theta^+
                     + \phi^- \theta^+
                     + 3 \, (\theta^+ \theta^-) \lambda_T^{+-}
                     - (\theta^+ \gamma^m \theta^-) \gamma_m \lambda_T^{+-}
\nonumber \\ & &
                     + (\theta^+ \theta^+) \lambda_T^{--}
      + (\theta^+ \gamma^m \theta^-) \chi_m^{(+)}
      + (\theta^+ \theta^+) \Zminusslash  \theta^- \,,
\end{eqnarray}
where the vector--spinor field $\chi^m$ is expressed in terms of
the spin-$\ft32$ field with symmetrized spinor indices in the following way
\begin{equation}
\chi^{(+)m \alpha} =
\left( \gamma^m \right)_{\beta\gamma}\, \chi^{(+)(\alpha\beta\gamma)}
\label{chispin32}
\end{equation}
and where, as usual, $\Aplusslash = \gamma^m \, A_m$.
\par
The superfield $\Phi_{gravitino}$
is linear in  the sense that it does not depend
on the monomial $\theta^-\theta^-$, but to be precise it is
a  spinorial superfield (\ref{4Dsuperfield})
with hypercharge $y_0$
that fulfils the stronger constraint
\begin{equation}
{\cal D}^+_\alpha \Phi_{gravitino}^\alpha = 0 \,.
\label{lineargravitino}
\end{equation}
\par
The generic  linear spinor superfield
contains, in its expansion, also terms  of the form
$\varphi^+ \theta^-$ and $(\theta^+\theta^+) \varphi^-\theta^-$,
where $\varphi^+$ and $\varphi^-$ are
 scalar fields and a term $(\theta^+ \g^m \theta^-) \chi_m$
where the spinor-vector $\chi_m$ is not an irreducible $\ft{3}{2}$
representation since it cannot be written as in (\ref{chispin32}).
 \par
Explicitly we have:
\begin{eqnarray}
\Phi^\alpha_{linear} &=& \lambda_L + \Aplusslash \theta^- + \Aminusslash \theta^+
                     + \phi^- \theta^+ + \varphi^+ \theta^-
                     + 3 \, (\theta^+ \theta^-) \lambda_T^{+-}
                     + (\theta^+ \theta^+) \lambda_T^{--}
\nonumber \\ & &
      + (\theta^+ \gamma^m \theta^-) \chi_m
      + (\theta^+ \theta^+) \Zminusslash  \theta^- + (\theta^+\theta^+) \varphi^-\theta^-\,,
\end{eqnarray}
The  field component $\chi^{\alpha m}$ in a generic unconstrained
spinor superfield can be decomposed in
a spin-$\ft12$ component and a spin-$\ft32$ component according to,
\begin{eqnarray}
\begin{array}{|c|c|}
\hline
\,\,\,   & \,\,\,  \cr
\hline
\end{array}
\times
\begin{array}{|c|}
\hline
\,\,\,  \cr
\hline
\end{array}
=
\begin{array}{c}
\begin{array}{|c|c|}
\hline
\,\,\,   & \,\,\,  \cr
\hline
\end{array}
\cr
\begin{array}{|c|}
\,\,\, \cr
\hline
\end{array}
\begin{array}{c}
\,\,\, \cr
\end{array}
\end{array}
+
\begin{array}{|c|c|c|}
\hline
\,\,\,   & \,\,\,  & \,\,\, \cr
\hline
\end{array}
\label{gravitinodecomposition}
\end{eqnarray}
where $m={ \begin{array}{|c|c|}
\hline
  &   \cr
\hline
\end{array}}\, $.
Then the constraint (\ref{lineargravitino}) eliminates
the scalars $\varphi^\pm$ and eliminates
the ${  \begin{array}{c}
\begin{array}{|c|c|}
\hline
   &   \cr
\hline
\end{array}
\cr
\begin{array}{|c|}
 \cr
\hline
\end{array}
\begin{array}{c}
 \cr
\end{array}
\end{array}}$-component of $\chi$ in terms of $\lambda_T^{+-}$.
From
\begin{equation}
s_\beta^-\, \Phi_{gravitino}^\alpha\Big\vert_{\theta^-\theta^-}
= \ft12 \left( -D_0 - y_0 - \ft32 \right)(\theta^- \theta^-)
     (\Aplusslash)_\beta{}^\alpha
\end{equation}
we conclude that the constraint (\ref{lineargravitino}) is superconformal
invariant if and only if
\begin{equation}
D_0 = - y_0 - \ft32 \,.
\end{equation}
\par
Once again we have retrieved the shortening condition already known
in the $SO(3) \times SO(2)$ basis: $E_0 =   \vert y_0 \vert + \ft32$
\subsubsection{Superfield description of the short graviton
multiplet}
For the massive short graviton multiplet we have the following states
\begin{eqnarray}
\centering
\begin{array}{||c|c|c||}
\hline
{\rm  Spin} &
{\rm Particle \, \, states}     &
{\rm Name}  \\
\hline
\hline
 & &  \\
2      & \bar a^-_{(\alpha} \bar a^+_\beta
\vert E_0=y_0+2, s_0=1,  y_0 \rangle_{\gamma\delta)}  & h \\
\ft32  & (\bar a^- \bar a^-) \bar a^+_{(\a} \vert y_0+2, 1,  y_0
\rangle_{\b\g)} & \chi^{(-)} \\
\ft32  & \bar a^+_{(\a} \vert y_0+2, 1,  y_0 \rangle_{\b\g)}  & \chi^{(+)}  \\
\ft32  & \bar a^-_{(\a} \vert y_0+2, 1,  y_0 \rangle_{\b\g)}  & \chi^{(+)} \\
1      & ( \bar a^- \bar a^- ) \vert y_0+2, 1,  y_0 \rangle  & Z \\
1      & ( \bar a^+ \bar a^- )  \vert y_0+2, 1,  y_0 \rangle   & Z \\
1      &   \vert E_0=y_0+2, 1,  y_0 \rangle       & A \\
\ft12  & \bar a^- \tau\cdot  \vert E_0=y_0+2, 1,  y_0 \rangle  & \lambda_T \\
 & &  \\
\hline
\hline
 & &  \\
2      & \bar a^-_{(\alpha} \bar a^+_\beta
\vert E_0=y_0+2, s_0=1,  - y_0 \rangle_{\gamma\delta)}  & h \\
\ft32  & (\bar a^+ \bar a^+) \bar a^-_{(\a} \vert y_0+2, 1,  - y_0
\rangle_{\b\g)} & \chi^{(-)} \\
\ft32  & \bar a^-_{(\a} \vert y_0+2, 1,  - y_0 \rangle_{\b\g)}
& \chi^{(+)}  \\
\ft32  & \bar a^+_{(\a} \vert y_0+2, 1,  - y_0 \rangle_{\b\g)}
& \chi^{(+)} \\
1      & ( \bar a^+ \bar a^+ ) \vert y_0+2, 1,  - y_0 \rangle  & Z \\
1      & ( \bar a^+ \bar a^- )  \vert y_0+2, 1,  - y_0 \rangle   & Z \\
1      &   \vert E_0=y_0+2, 1,  - y_0 \rangle       & A \\
\ft12  & \bar a^+ \tau\cdot  \vert E_0=y_0+2, 1,  - y_0 \rangle  & \lambda_T \\
 & &  \\
\hline
\end{array}
\end{eqnarray}
Applying the rotation $U$ (\ref{rotationmatrix}) to the upper part of the above table,
and identifying the particle states with the corresponding boundary fields, as
we have done so far, we derive the short graviton superfield:
\begin{eqnarray}
\Phi^m_{graviton} &=&
A^m + \theta^+ \gamma^m \lambda_T^-
+  \theta^- \chi^{(+)+m} + \theta^+ \chi^{(+)-m}
\nonumber \\
& &
+ (\theta^+ \theta^-) \, Z^{+-m}
+ \ft{i}{2}\, \varepsilon^{mnp} \, ( \theta^+ \gamma_n \theta^-) \,  Z_p^{+-}
+ (\theta^+ \theta^+) \, Z^{--m}
\nonumber \\
& &
+ ( \theta^+ \gamma_n \theta^-) \, h^{mn} +
(\theta^+ \theta^+) \, \theta^- \chi^{(-)-m}
\,,
\end{eqnarray}
where
\begin{eqnarray}
\chi^{(+)\pm m \alpha} &=&
\left( \g^m \right)_{\beta\gamma}\chi^{(+)\pm(\alpha\beta\gamma)} \,,
\nonumber \\
\chi^{(-)- m \alpha} &=&
\left( \g^m \right)_{\beta\gamma}\chi^{(-)-(\alpha\beta\gamma)} \,,
\nonumber \\
h^m{}_m &=& 0 \,,
\label{gravitonchispin32}
\end{eqnarray}
This superfield  satisfies the following constraint,
\begin{eqnarray}
{\cal D}^+_\alpha \Phi_{graviton}^{\alpha \beta}=0 \,,
\end{eqnarray}
where we have defined:
\begin{eqnarray}
\Phi^{\alpha\beta}= \left(\gamma_m\right)^{\alpha\beta} \, \Phi^m
\end{eqnarray}
Furthermore we check that $s^- \Phi^m_{graviton}$ is still a short graviton
superfield if and only if:
\begin{equation}
D_0 = - y_0 - 2 \,.
\end{equation}
corresponding to the known unitarity bound \cite{multanna,m111spectrum}:
\begin{equation}
  E_0 = \vert y_0 \vert +2
\label{graunibou}
\end{equation}
\subsubsection{Superfield description of the massless vector
multiplet}
\par
Considering now ultrashort multiplets we focus on the massless vector
multiplet containing the following bulk particle states:
\begin{eqnarray}
\centering
\begin{array}{||c|c|c||}
\hline
{\rm  Spin} &
{\rm Particle \, \, states}     &
{\rm Name}  \\
\hline
\hline
 & &  \\
1     &  \bar a^-_1 \bar a^+_1
\vert E_0=1 , s_0=0,  y_0=0 \rangle \,,
\bar a^-_2 \bar a^+_2
\vert E_0=1 , s_0=0,  y_0=0 \rangle\,\,  & A \\
\ft12  & \bar a^+ \vert 1 , 0, 0  \rangle
 & \lambda_L \\
\ft12  & \bar a^- \vert 1 , 0, 0  \rangle
 & \lambda_L \\
0  &  \bar a^- \bar a^+ \vert 1, 0,  0 \rangle  & \pi \\
0      & \vert 1, 0,  0 \rangle  & S \\
 & &  \\
\hline
\end{array}
\end{eqnarray}
where the gauge field  $A$ has only two  helicity states
$1$ and $-1$.
Applying the rotation $U$ (\ref{rotationmatrix}) we get,
\begin{eqnarray}
V= S + \theta^+ \lambda_L^- + \theta^- \lambda_L^+
+ (\theta^+ \theta^-) \, \pi +   \theta^+ \Aslash \theta^- \,.
\label{Vfixed}
\end{eqnarray}
This multiplet can be obtained by a real superfield
\begin{eqnarray}
V &=& S + \theta^+ \lambda_L^- + \theta^- \lambda_L^+
+ (\theta^+ \theta^-) \, \pi +   \theta^+ \Aslash \theta^-
\nonumber \\ & &
+ (\theta^+\theta^+) \, M^{--} + (\theta^-\theta^-) \, M^{++}
\nonumber \\
& &
+ (\theta^+ \theta^+) \, \theta^- \mu^-
+ (\theta^- \theta^-) \, \theta^+ \mu^+
\nonumber \\ & &
+ (\theta^+\theta^+) (\theta^-\theta^-) \, F \,, \nonumber\\
V^\dagger &=& V
\end{eqnarray}
that transforms as follows under a gauge transformation\footnote{The vector component
transforms under a $SU(2)$
or a $SU(3)$ gauge transformation
in the case of \cite{m111spectrum}.} ,
\begin{eqnarray}
V \rightarrow V + \Lambda + \Lambda^\dagger \,,
\end{eqnarray}
where $\Lambda$ is a chiral superfield of the form
(\ref{formchiral}). In components
this reads,
\begin{eqnarray}
S &\rightarrow& S + X + X^* \,, \nonumber\\
\lambda_L^- &\rightarrow& \lambda_L^- + \lambda \,, \nonumber \\
\pi &\rightarrow& \pi \,, \nonumber \\
A_m &\rightarrow& A_m + \ft{1}{2}\, \partial_m\left(X-X^*\right) \,,
\nonumber \\
M^{--} &\rightarrow& M^{--} + H \,, \nonumber \\
\mu^{-} &\rightarrow& \mu^- +\ft{1}{4} \, \dslash\lambda \,, \nonumber \\
F &\rightarrow& F +\ft{1}{16} \, \Box  X \,,
\end{eqnarray}
which may be used to gauge fix the real multiplet in the
following way,
\begin{equation}
M^{--}= M^{++}=\mu^-=\mu^+= F =0\,,
\label{gaugefixingvector}
\end{equation}
to obtain (\ref{Vfixed}).
For the scaling weight $D_0$ of the massless vector multiplet
we find $-1$. Indeed this follows from the fact that $\Lambda$
is a chiral superfield with $y_0=0, D_0=0$. Which is also in
agreement with $E_0=1$ known from \cite{multanna, m111spectrum}.
\par
\subsubsection{Superfield description of the massless graviton
multiplet}
The massless graviton multiplet is composed of the following bulk
particle states:
\begin{eqnarray}
\centering
\begin{array}{||c|c|c||}
\hline
{\rm  Spin} &
{\rm Particle \, \, states}     &
{\rm Name}  \\
\hline
\hline
 & &  \\
2     &  \bar a^- \tau^{(i} \bar a^+
\vert E_0=2 , s_0=1,  y_0=0 \rangle^{j)}  & h \\
\ft32  & \bar a^+ \vert 2 , 1, 0  \rangle^i
 & \chi^{(+)} \\
\ft32  & \bar a^- \vert 2 , 1, 0  \rangle^i
 & \chi^{(+)} \\
1  &  \vert 2, 1,  0 \rangle^i  & A \\
 & &  \\
\hline
\end{array}
\end{eqnarray}
from which, with the usual procedure we obtain
\begin{eqnarray}
g_m = A_m + \theta^+ \chi^{(+)-}_m
    + \theta^- \chi^{(+)+}_m +  \theta^+ \gamma^n \theta^-\, h_{mn}\,.
\end{eqnarray}
Similarly as for the vector multiplet we may write this
multiplet as a gauge fixed multiplet with local gauge symmetries
that include local coordinate transformations, local supersymmetry
and local $SO(2)$, in other words full supergravity. However this is not the
goal of our work where we prepare to interprete the bulk gauge fields as composite states
in the boundary conformal field theory..
\par
This completes the treatment of the short $Osp(2\vert 4)$
multiplets of \cite{m111spectrum}. We have found that
all of them are linear multiplets with the extra constraint
that they have to transform into superfields of the
same type under $s$-supersymmetry. Such constraint is identical with
the shortening conditions found in the other constructions of unitary
irreducible representations of the orthosymplectic superalgebra.
%
%
\section{${\cal N}=2$, $d=3$ gauge theories
and their rheonomic construction}
\label{n2d3gauge}
Next, as announced in the introduction, we turn to consider gauge
theories in three space--time dimensions with ${\cal N}=2$
supersymmetry. From the view point of the $AdS_4/CFT_3$
correspondence these gauge theories, whose elementary fields we
collectively denote $\phi^i(x)$, are microscopic field theories
living on the $M2$ brane world volume such that suitable composite operators
(see also fig.\ref{gaugefig}):
\begin{equation}
  {\cal O}(x) = \phi^{i_1} (x, ) \, \dots \, \phi^{i_n}(x)
  c_{i_1 \dots i_n}
\label{marmellata}
\end{equation}
\iffigs
\begin{figure}
\caption{Conformal superfields are composite operators in the gauge theory
\label{gaugefig}}
\begin{center}
\epsfxsize = 13cm
\epsffile{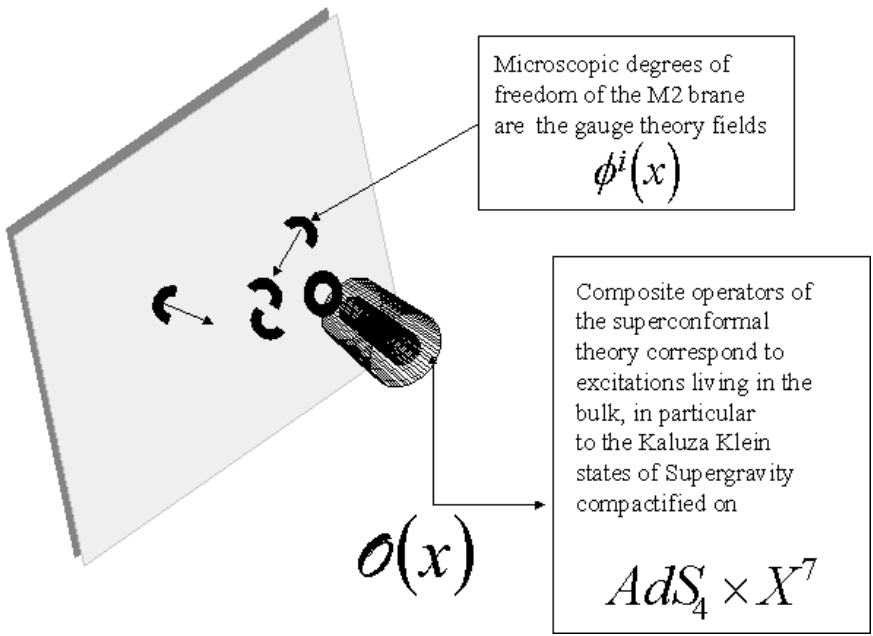}
\vskip  0.2cm
\hskip 2cm
\unitlength=1.1mm
\end{center}
\end{figure}
\fi
\par
\par
can be identified with the components of the conformal superfields
described in the previous section and matching the Kaluza Klein
classical spectrum.
\par
According to the specific horizon $X^7$, the world volume gauge group is of
the form:
\begin{equation}
  G^{WV}_{gauge}= U(k_1 \,N)^{\ell_1} \, \times \, \dots \,\times \, U(k_n
  N)^{\ell_n}
\label{Wvgauge}
\end{equation}
where $k_i$ and $\ell_i$ are integer number and where the
correspondence is true in the large $N \to \infty$ limit. Indeed $N$
is to be identified with the number of $M2$ branes in the stack.
\par
In addition the gauge theory has a {\it flavor group} which coincides
with the gauge group of Kaluza Klein supergravity, namely with the
isometry group of the $X^7$ horizon:
\begin{equation}
  G^{WV}_{flavor} = G_{KK}^{bulk} = \mbox{isometry}(X^7)
\label{flavgrou}
\end{equation}
Since our goal is to study the general features of the
$AdS_4/CFT_3$ correspondence, rather than specific cases, we concentrate on the
construction of a generic ${\cal N}=2$ gauge theory with an arbitrary
gauge group and an arbitrary number of chiral multiplets in generic
interaction. We are mostly interested in the final formulae for the
scalar potential   and on the restrictions that guarantee
an enlargement of supersymmetry to ${\cal N}=4$ or ${\cal N}=8$,
but we provide a complete construction of the lagrangian and of the
supersymmetry transformation rules. To this effect we utilize the
time honored method of rheonomy \cite{castdauriafre,billofre,fresoriani}
that yields the the result for the lagrangian and the supersymmetry rules
in component form avoiding the too much implicit notation of superfield
formulation.
The first step in the rheonomic construction of a rigid supersymmetric theory
involves writing the structural equations of rigid superspace.
\subsection{${\cal N}=2, \, d=3$ rigid superspace}
The $d\!\!=\!\!3,~{\cal N}$--extended superspace is
viewed as the supercoset space:
\begin{equation}
{\cal M}^{\cal N}_3=\frac{ISO(1,2|{\cal N})}
{SO(1,2) }\, \equiv \, \frac{Z\left[ ISO(1,2|{\cal N})\right ]
}{SO(1,2) \times \IR^{{\cal N}({\cal N}-1)/2}}
\end{equation}
where $ISO(1,2\vert{\cal N})$ is the ${\cal N}$--extended Poincar\'e
superalgebra in three--dimensions. It is the subalgebra of
$Osp({\cal N}\vert 4)$ (see eq. (\ref{ospD})) spanned by the generators
$J_m$, $P_m$, $q^i$.
The central extension $Z\left[ ISO(1,2|{\cal N})\right ]$ which is
not contained in $Osp({\cal N}\vert 4)$ is obtained by adjoining to
$ISO(1,2|{\cal N})$ the central charges that generate the subalgebra
$\IR^{{\cal N}({\cal N}-1)/2}$.
Specializing our analysis to the case ${\cal N}\!\!=\!\!2$, we can
define the new generators:
\begin{equation}
\left\{\begin{array}{ccl}
Q&=&\sqrt{2}q^-=(q^1-iq^2)\\
Q^c&=&\sqrt{2}iq^+=i(q^1+iq^2)\\
Z&=&Z^{12}
\end{array}\right.
\end{equation}
The left invariant one--form $\Omega$ on ${\cal M}^{\cal N}_3$ is:
\begin{equation}
\Omega=iV^mP_m-i\o^{mn}J_{mn}+i\overline{\psi^c}Q-i\overline{\psi}Q^c
+i{\cal B}Z\,.
\end{equation}
The superalgebra (\ref{ospD}) defines  all the structure constants
apart from those relative to the central charge that are trivially
determined. Hence we can write:
\begin{eqnarray}
d\Omega-\Omega\wedge\Omega&=&\left(dV^m-\o^m_{\ n}\wedge V^n
+i\overline{\psi}\wedge\g^m\psi
+i\overline{\psi}^c\wedge\g^m\psi^c\right)P_m\nonumber\\
&&-\ft{1}{2}i\left(d\o^{mn}-\o^m_{\ p}\wedge\o^{pn}\right)J_{mn}\nonumber\\
&&+\left(d\overline{\psi}^c
+\ft{1}{2}\o^{mn}\wedge\overline{\psi}^c\g_{mn}\right)Q\nonumber\\
&&+\left(d\overline{\psi}
-\ft{1}{2}\o^{mn}\wedge\overline{\psi}\g_{mn}\right)Q^c\nonumber\\
&&+i\left(d{\cal B}+i\overline{\psi}^c\wedge\psi^c
-i\overline{\psi}\wedge\psi\right)Z
\end{eqnarray}
Imposing the Maurer-Cartan equation $d\Omega-\Omega\wedge\Omega=0$
is equivalent to imposing flatness in superspace, i.e. global
supersymmetry. So we have
\begin{equation}
\left\{\begin{array}{ccl}
dV^m-\o^m_{\ n}\wedge V^n&=&-i\overline{\psi}^c\wedge\g^m\psi^c
-i\overline{\psi}\wedge\g^m\psi\\
d\o^{mn}&=&\o^m_{\ p}\wedge\o^{pn}\\
d\overline{\psi}^c&=&-\ft{1}{2}\o^{mn}\wedge\overline{\psi}^c\g_{mn}\\
d\overline{\psi}&=&\ft{1}{2}\o^{mn}\wedge\overline{\psi}\g_{mn}\\
d{\cal B}&=&-i\overline{\psi}^c\wedge\psi^c
+i\overline{\psi}\wedge\psi
\end{array}\right.
\end{equation}
The simplest solution for the supervielbein and connection is:
\begin{equation}
\left\{\begin{array}{ccl}
V^m&=&dx^m-i\overline{\theta}^c\g^md\theta^c
-i\overline{\theta}\g^m d\theta\\
\omega^{mn}&=&0\nonumber\\
\psi&=&d\theta\\
\psi^c&=&d\theta^c\\
{\cal B}&=&-i\overline{\theta}^c\,d\theta^c
+i\overline{\theta}\,d\theta
\end{array}\right.
\end{equation}
The superderivatives discussed in the previous
sections (compare with \eqn{supdervcf}),
\begin{equation}
\left\{\begin{array}{ccl}
D_m&=&\partial_m\\
D&=&\frac{\partial}{\partial\overline{\theta}}-i\g^m\theta\partial_m\\
D^c&=&\frac{\partial}{\partial\overline{\theta}^c}-i\g^m\theta^c\partial_m\\
\end{array}\right.,
\end{equation}
are the vectors dual to these one--forms.
%
%
\subsection{Rheonomic construction of the ${\cal N}=2,~d=3,$ lagrangian}
As stated we are interested in the generic form of ${\cal N}=2,~d=3$
super Yang Mills theory coupled to $n$ chiral multiplets arranged into
a generic representation $\cal R$ of the gauge group $\cal G$.
\par
In ${\cal N}=2,~d=3$ supersymmetric theories, two formulations
are allowed: the on--shell and the off--shell one.
In the on--shell formulation which contains only
the physical fields, the supersymmetry transformations rules
close the supersymmetry algebra only
upon use of  the field equations.
On the other hand the off--shell formulation contains further auxiliary, non
dynamical fields that make it possible for
 the supersymmetry transformations rules to close the
supersymmetry algebra identically.
By solving the field equations of the auxiliary fields
these latter can be eliminated   and the on--shell formulation can be retrieved.
We  adopt the off--shell formulation.
%
%
\subsubsection{The gauge multiplet}
The three--dimensional ${\cal N}=2$ vector multiplet contains the following
Lie-algebra valued fields:
\begin{equation}\label{vectorm}
\left({\cal A},\l,\l^c,M,P\right)\,,
\end{equation}
where ${\cal A}={\cal A}^It_I$ is the real gauge connection one--form,
$\l$ and $\l^c$ are two complex Dirac spinors (the \emph{gauginos}),
$M$ and $P$ are real scalars; $P$ is an auxiliary field.
\par
The field strength is:
\begin{equation}\label{defF}
F=d{\cal A}+i{\cal A}\wedge{\cal A}\,.
\end{equation}
The covariant derivative on the other fields of the gauge multiplets
is defined as:
\begin{equation}\label{defnablagauge}
\nabla X=dX+i\left[{\cal A},X\right]\,.
\end{equation}
From (\ref{defF}) and (\ref{defnablagauge}) we obtain the Bianchi identity:
\begin{equation}
\nabla^2 X=i\left[F,X\right]\,.
\end{equation}
The rheonomic parametrization of the \emph{curvatures} is given by:
\begin{equation}\label{vectorB}
\left\{\begin{array}{ccl}
F&=&F_{mn}V^mV^n-i\overline{\psi}^c\g_m\l V^m-i\overline{\psi}\g_m\l^cV^m
+iM\left(\overline{\psi}\psi-\overline{\psi}^c\psi^c\right)\\
\nabla\l&=&V^m\nabla_m\l+\nslash M\psi^c-F_{mn}\g^{mn}\psi^c
+iP\psi^c\\
\nabla\l^c&=&V^m\nabla_m\l^c-\nslash M\psi-F_{mn}\g^{mn}\psi
-iP\psi\\
\nabla M&=&V^m\nabla_mM+i\overline{\psi}\l^c-i\overline{\psi}^c\l\\
\nabla P&=&V^m\nabla_mP+\overline{\psi}\nslash\l^c-\overline{\psi}^c\nslash\l
-i\overline{\psi}\left[\l^c,M\right]-i\overline{\psi}^c\left[\l,M\right]
\end{array}\right.
\end{equation}
and we also have:
\begin{equation}
\left\{\begin{array}{ccl}
\nabla F_{mn}&=&V^p\nabla_pF_{mn}+i\overline{\psi}^c\g_{[m}\nabla_{n]}\l
+i\overline{\psi}\g_{[m}\nabla_{n]}\l^c\\
\nabla\nabla_mM&=&V^n\nabla_n\nabla_mM+i\overline{\psi}\nabla_m\l^c
-i\overline{\psi}^c\nabla_m\l+\overline{\psi}^c\g_m\left[\l,M\right]
+\overline{\psi}\g_m\left[\l^c,M\right]\\
\nabla\nabla_m\l&=&V^n\nabla_n\nabla_m\l+\nabla_m\nabla_nM\g^n\psi^c
-\nabla_mF_{np}\g^{np}\psi^c\\
&&+i\nabla_mP\psi^c+\overline{\psi}\g_m\left[\l^c,\l\right]\\
\nabla_{[p}F_{mn]}&=&0\\
\nabla_{[m}\nabla_{n]}M&=&i\left[F_{mn},M\right]\\
\nabla_{[m}\nabla_{n]}\l&=&i\left[F_{mn},\l\right]
\end{array}\right.
\end{equation}
The off--shell formulation of the theory contains an arbitrariness
in the choice of the functional dependence of the auxiliary fields
on the physical fields.
Consistency with the Bianchi identities forces the generic expression of $P$ as a function
of $M$ to be:
\begin{equation}\label{defP}
P^I=2\a M^I+\zeta^{\tilde I}{\cal C}_{\tilde I}^{\ I}\,,
\end{equation}
where $\a,~\zeta^{\tilde I}$ are arbitrary real parameters and
${\cal C}_{\tilde I}^{\ I}$ is the projector on the center of
$Z[\cal G]$ of the gauge Lie algebra.
The terms in the lagrangian proportional to $\a$ and $\zeta$ are separately
supersymmetric.
In the bosonic lagrangian, the part proportional to $\a$ is a Chern Simons
term, while the part proportional to $\zeta$ constitutes the Fayet Iliopoulos
term.
Note that the Fayet Iliopoulos terms are associated only with a
central abelian subalgebra of the gauge algebra $\cal G$.
\par
Enforcing (\ref{defP}) we get the following equations of motion for the spinors:
\begin{equation}
\left\{\begin{array}{c}
\nslash\l=2i\a\l-i\left[\l,M\right]\\
\\
\nslash\l^c=2i\a\l^c+i\left[\l^c,M\right]
\end{array}\right.
\end{equation}
Taking the covariant derivatives of these, we obtain the equations of motion
for the bosonic fields:
\begin{equation}
\left\{\begin{array}{c}
\nabla_m\nabla^m M^I=-4\a^2M^I-2\a\zeta^{\tilde I}{\cal C}_{\tilde I}^{\ I}
-2\left[\overline{\l},\l\right]^I\\
\nabla^n F_{mn}=-\a\e_{mnp}F^{np}-\ft{i}{2}\left[\nabla_mM,M\right]
\end{array}\right.
\end{equation}
Using the rheonomic approach we find the following superspace lagrangian
for the gauge multiplet:
\begin{equation}\label{gaugeLag}
{\cal L}_{gauge}={\cal L}_{gauge}^{Maxwell}
+{\cal L}_{gauge}^{Chern-Simons}
+{\cal L}_{gauge}^{Fayet-Iliopoulos}\,,
\end{equation}
where
\begin{eqnarray}
{\cal L}_{gauge}^{Maxwell}&=&Tr\left\{-F^{mn}\left[F
+i\overline{\psi}^c\g_m\l V^m+i\overline{\psi}\g_m\l^cV^m
-2iM\overline{\psi}\psi\right]V^p\epsilon_{mnp}\right.\nonumber\\
&+&\ft{1}{6}F_{qr}F^{qr}V^m V^n V^p\epsilon_{mnp}
-\ft{1}{4}i\epsilon_{mnp}\left[\nabla\overline{\l}\g^m\l
+\nabla\overline{\l}^c\g^m\l^c\right]
V^n V^p\nonumber\\
&+&\ft{1}{2}\epsilon_{mnp}{\cal M}^m\left[\nabla M-i\overline{\psi}\l^c
+i\overline{\psi}^c\l\right] V^nV^p
-\ft{1}{12}{\cal M}^d{\cal M}_d\epsilon_{mnp}V^mV^nV^p\nonumber\\
&+&\nabla M \overline{\psi}^c\g_c\l V^p
-\nabla M \overline{\psi}\g_p\l^cV^p\nonumber\\
&+&F\overline{\psi}^c\l+F\overline{\psi}\l^c
+\ft{1}{2}i\overline{\l}^c\l\overline{\psi}^c\g_m\psi V^m
+\ft{1}{2}i\overline{\l}\l^c\overline{\psi}\g_m\psi^cV^m\nonumber\\
&+&\ft{1}{12}{\cal P}^2V^mV^nV^p\epsilon_{mnp}
-2i(\overline{\psi}\psi)M\left[\overline{\psi}^c\l
+\overline{\psi}\l^c\right]\nonumber\\
&-&\left.\ft{1}{6}M\left[\overline\l,\l\right]V^mV^nV^p\e_{mnp}\right\}\,,
\end{eqnarray}
\begin{eqnarray}
{\cal L}_{gauge}^{Chern-Simons}&=&\a Tr\left\{
-\left(A\wedge F-iA\wedge A\wedge A\right)
-\ft{1}{3}MP\epsilon_{mnp}V^m V^n V^p\right.\nonumber\\
&+&\ft{1}{3}\overline{\l}\l\epsilon_{mnp}V^m V^n V^p
+M\epsilon_{mnp}\left[\overline{\psi}^c\g^m\l
-\overline{\psi}\g^m\l^c\right]V^n V^p\nonumber\\
&&\left.-2iM^2\overline{\psi}\g_m\psi V^m\right\}
\end{eqnarray}
\begin{eqnarray}
{\cal L}_{gauge}^{Fayet-Iliopoulos}&=&Tr\left\{\zeta{\cal C}\left[
-\ft{1}{6}P\e_{mnp}V^mV^nV^p
+\ft{1}{2}\e_{mnp}\left(\overline{\psi}^c\g^m\l
-\overline{\psi}\g^m\l^c\right)V^nV^p
\right.\right.\nonumber\\
&&\left.\left.-2iM\overline{\psi}\g_m\psi V^m
-2i{\cal A}\overline{\psi}\psi\right]\right\}
\end{eqnarray}
%
%
\subsubsection{Chiral multiplet}
The chiral multiplet contains the following fields:
\begin{equation}\label{chiralm}
\left(z^i,\chi^i,H^i\right)
\end{equation}
where $z^i$ are complex scalar fields which parametrize a  K\"ahler
manifold. Since we are interested in microscopic theories
with canonical kinetic terms we take this K\"ahler manifold to be
flat and we choose its metric to be the constant
$\eta_{ij^*}\equiv \mbox{diag}(+,+,\dots,+)$. The other fields in
the chiral multiplet are $\chi^i$ which is a two components Dirac spinor
and $H^i$ which is a complex scalar auxiliary field.
The index $i$ runs in the representation $\cal R$ of $\cal G$.
\par
The covariant derivative of the fields $X^i$ in the chiral multiplet is:
\begin{equation}
\nabla X^i=dX^i+i\eta^{ii^*}{\cal A}^I(T_I)_{i^*j}X^j\,,
\end{equation}
where $(T_I)_{i^*j}$ are the hermitian generators of $\cal G$
in the representation $\cal R$.
The covariant derivative of the complex conjugate fields
$\overline X^{i^*}$ is:
\begin{equation}
\nabla\overline X^{i^*}=d\overline X^{i^*}
-i\eta^{i^*i}{\cal A}^I(\overline T_I)_{ij^*}\overline X^{j^*}\,,
\end{equation}
where
\begin{equation}
(\overline T_I)_{ij^*}\equiv\overline{(T_I)_{i^*j}}=(T_I)_{j^*i}\,.
\end{equation}
The rheonomic parametrization of the curvatures is given by:
\begin{equation}\label{chiralB}
\left\{\begin{array}{ccl}
\nabla z^i&=&V^m\nabla_m z^i+2\overline{\psi}^c\chi^i\\
\nabla\chi^i&=&V^m\nabla_m\chi^i-i\nslash z^i\psi^c+H^i\psi
-M^I(T_I)^i_{\,j}z^j\psi^c\\
\nabla H^i&=&V^m\nabla_m H^i-2i\overline{\psi}\nslash\chi^i
-2i\overline{\psi}\l^I(T_I)^i_{\,j} z^j
+2M^I(T_I)^i_{\,j}\overline{\psi}\chi^j
\end{array}\right.\,.
\end{equation}
We can choose the auxiliary fields $H^i$ to be the derivatives of an
arbitrary antiholomorphic superpotential $\overline W(\overline z)$:
\begin{equation}\label{defW^*}
H^i=\eta^{ij^*}\frac{\partial\overline W(\overline z)}{\partial z^{j^*}}
=\eta^{ij^*}\partial_{j^*}\overline W
\end{equation}
Enforcing eq. (\ref{defW^*}) we get the following equations of motion
for the spinors:
\begin{equation}\label{chimotion}
\left\{\begin{array}{c}
\nslash\chi^i=i\eta^{ij^*}\partial_{j^*}\partial_{k^*}
\overline W\chi^{ck^*}-\l^I(T_I)^i_{\,j} z^j-iM^I(T_I)^i_{\,j}\chi^j\\
\\
\nslash\chi^{ci^*}=i\eta^{i^*j}\partial_j\partial_k W\chi^k
+\l^{cI}(\overline T_I)^{i^*}_{\,j^*}\overline z^{j^*}
-iM^I(\overline T_I)^{i^*}_{\,j^*}\chi^{cj^*}
\end{array}\right.\,.
\end{equation}
Taking the differential of (\ref{chimotion}) one obtains the equation of
motion for $z$:
\begin{eqnarray}
\Box z^i&=&\eta^{ii^*}\partial_{i^*}\partial_{j^*}\partial_{k^*}
\overline W(\overline z)\left(\overline{\chi}^{j^*}\chi^{ck^*}\right)
-\eta^{ij^*}\partial_{j^*}\partial_{k^*}
\overline W(\overline z)\partial_i W(z)\nonumber\\
&&+P^I(T_I)^i_{\,j}z^j-M^IM^J(T_IT_J)^i_{\,j}z^j
-2i\overline{\l}^I(T_I)^i_{\,j}\chi^j
\end{eqnarray}
%
%
The first order Lagrangian for the chiral multiplet (\ref{chiralm}) is:
\begin{equation}\label{chiralLag}
{\cal L}_{chiral}={\cal L}_{chiral}^{Wess-Zumino}
+{\cal L}_{chiral}^{superpotential}\,,
\end{equation}
where
\begin{eqnarray}
{\cal L}_{chiral}^{Wess-Zumino}&=&\ft{1}{2}\e_{mnp}\overline{\Pi}^{m\,i^*}
\eta_{i^*j}\left[\nabla z^j
-2\overline{\psi}^c\chi^j\right]V^nV^p\nonumber\\
&+&\ft{1}{2}\e_{mnp}\Pi^{m\,i}\eta_{ij^*}\left[\nabla \overline z^{j^*}
-2\overline{\chi}\psi^{c\,j^*}\right]V^nV^p\nonumber\\
&-&\ft{1}{6}\e_{mnp}\eta_{ij^*}\Pi_q^{\,i}
\overline{\Pi}^{q\,j^*}V^mV^nV^p\nonumber\\
&+&\ft{1}{2}i\e_{mnp}\eta_{ij^*}\left[\overline{\chi}^{j^*}\g^m\nabla\chi^i
+\overline{\chi}^{c\,i}\g^m\nabla\chi^{c\,j^*}
\right]V^nV^p\nonumber\\
&+&2i\eta_{ij^*}\left[\nabla z^i\overline{\psi}\g_m\chi^{c\,j^*}
-\nabla \overline z^{j^*}\overline{\chi}^{c\,i}\g_m\psi
\right]V^m\nonumber\\
&-&2i\eta_{ij^*}\left(\overline{\chi}^{j^*}\g_m\chi^i\right)
\left(\overline{\psi}^c\psi^c\right)V^m
-2i\eta_{ij^*}\left(\overline{\chi}^{j^*}\chi^i\right)
\left(\overline{\psi}^c\g_m\psi^c\right)V^m\nonumber\\
&+&\ft{1}{6}\eta_{ij^*}H^i\overline H^{j^*}\e_{mnp}V^mV^nV^p
+\left(\overline{\psi}\psi\right)
\eta_{ij^*}\left[\overline z^{j^*}\nabla z^i
-z^i\nabla \overline z^{j^*}\right]\nonumber\\
&+&i\e_{mnp}z^iM^I(T_I)_{ij^*}\overline{\chi}^{j^*}\g^m\psi^c
V^nV^p\nonumber\\
&+&i\e_{mnp}\overline z^{j^*}M^I(T_I)_{j^*i}
\overline{\chi}^{c\,i}\g^m\psi V^nV^p\nonumber\\
&-&\ft{1}{3}M^I(T_I)_{ij^*}\overline{\chi}^{j^*}\chi^i
\e_{mnp}V^mV^nV^p\nonumber\\
&+&\ft{1}{3}i\left[\overline{\chi}^{j^*}\l^I(T_I)_{j^*i} z^i
-\overline{\chi}^{c\,i}\l^{c\,I}(T_I)_{ij^*}\overline z^{j^*}\right]
\e_{mnp}V^mV^nV^p\nonumber\\
&+&\ft{1}{6}z^iP^I(T_I)_{ij^*}\overline z^{j^*}\e_{mnp}V^mV^nV^p\nonumber\\
&-&\ft{1}{2}\left(\overline{\psi}^c\g^m\l^I(T_I)_{ij^*}\right)
z^i\overline z^{j^*}\e_{mnp}V^nV^p\nonumber\\
&+&\ft{1}{2}\left(\overline{\psi}\g^m\l^{c\,I}(T_I)_{ij^*}\right)
z^i\overline z^{j^*}\e_{mnp}V^nV^p\nonumber\\
&-&\ft{1}{6}M^IM^J\,z^i(T_IT_J)_{ij^*}\overline z^{j^*}
\e_{mnp}V^mV^nV^p\nonumber\\
&+&2iM^I(T_I)_{ij^*}z^i\overline z^{j^*}\overline{\psi}\g_m\psi V^m\,,
\end{eqnarray}
and
\begin{eqnarray}
{\cal L}_{chiral}^{superpotential}&=&
-i\e_{mnp}\left[\overline{\chi}^{j^*}\g^m\partial_{j^*}\overline W
(\overline z)\psi+\overline{\chi}^{c\,j}\g^m\partial_j\overline W(z)
\psi^c\right]V^nV^p\nonumber\\
&+&\ft{1}{6}\left[\partial_i\partial_jW(z)\overline{\chi}^{c\,i}\chi^j
+\partial_{i^*}\partial_{j^*}\overline W(\overline z)
\overline{\chi}^{i^*}\chi^{c\,j^*}\right]\e_{mnp}V^mV^nV^p\nonumber\\
&-&\ft{1}{6}\left[H^i\partial_iW(z)+\overline H^{j^*}\partial_{j^*}
\overline W(\overline z)\right]\e_{mnp}V^mV^nV^p\nonumber\\
&-&2i\left[W(z)+\overline W(\overline z)\right]
\overline{\psi}\g_m\psi^cV^m
\end{eqnarray}
%
%
\subsubsection{The space--time Lagrangian}
In the rheonomic approach (\cite{castdauriafre}), the total
three--dimensional ${\cal N}\!\!=\!\!2$ lagrangian:
\begin{equation}
{\cal L}^{{\cal N}=2}={\cal L}_{gauge}+{\cal L}_{chiral}
\end{equation}
is a closed ($d{\cal L}^{{\cal N}=2}=0$) three--form defined in
superspace.
The action is given by the integral of ${\cal L}^{{\cal N}=2}$ on
a generic \emph{bosonic} three--dimensional surface ${\cal M}_3$
in superspace:
\begin{equation}
S=\int_{{\cal M}_3}{\cal L}^{{\cal N}=2}\,.
\end{equation}
Supersymmetry transformations can be viewed as global translations in
superspace which move ${\cal M}_3$.
Then, being ${\cal L}^{{\cal N}=2}$ closed, the action is invariant under
global supersymmetry transformations.
\par
We choose as bosonic surface the one defined by:
\begin{equation}
\theta=d\theta=0\,.
\end{equation}
Then the space--time lagrangian, i.e. the pull--back of
${\cal L}^{{\cal N}=2}$ on ${\cal M}_3$, is:
\begin{equation}\label{N=2stLag}
{\cal L}^{{\cal N}=2}_{st}={\cal L}^{kinetic}_{st}
+{\cal L}^{fermion~mass}_{st}+{\cal L}^{potential}_{st}\,,
\end{equation}
where
\begin{eqnarray}\label{N=2chiralst}
\LL_{st}^{kinetic}&=&\left\{
\eta_{ij^*}\nabla_m z^i\nabla^m\overline z^{j^*}
+i\eta_{ij^*}\left(\overline{\chi}^{j^*}\nslash\chi^i
+\overline{\chi}^{c\,i}\nslash\chi^{c\,j^*}\right)
\right.\nonumber\\
&&-g_{IJ}F^I_{mn}F^{J\,mn}
+\ft{1}{2}g_{IJ}\nabla_m M^I\nabla^m M^J\nonumber\\
&&+\left.\ft{1}{2}ig_{IJ}\left(\overline{\l}^I\nslash\l^J
+\overline{\l}^{c\,I}\nslash\l^{c\,J}\right)\right\}d^3x\\
&&\nonumber\\
\LL_{st}^{fermion~mass}&=&
\left\{i\left(\overline{\chi}^{c\,i}\partial_i\partial_jW(z)\chi^j
+\overline{\chi}^{i^*}\partial_{i^*}\partial_{j^*}\overline W(\overline z)
\chi^{c\,j^*}\right)\right.\nonumber\\
&&-f_{IJK}M^I\overline{\l}^J\l^K
-2\overline{\chi}^{i^*}M^I(T_I)_{ij^*}\chi^{j^*}\nonumber\\
&&+2i\left(\overline{\chi}^{i^*}\l^I(T_I)_{i^*j}z^j
-\overline{\chi}^{c\,i}\l^I(T_I)_{ij^*}\overline z^{j^*}\right)\nonumber\\
&&\left.+2\a g_{IJ}\overline{\l}^I\l^J\right\}d^3x\\
\LL_{st}^{potential}&=&-U(z,\overline z,H,\overline H,M,P)d^3x\,,
\end{eqnarray}
and
\begin{eqnarray}\label{defU}
U(z,\overline z,H,\overline H,M,P)&=&
H^i\partial_iW(z)
+\overline H^{j^*}\partial_{j^*}\overline W(\overline z)
-\eta_{ij^*}H^i\overline H^{j^*}\nonumber\\
&&-\ft{1}{2}g_{IJ}P^IP^J-z^iP^I(T_I)_{ij^*}\overline z^{j^*}\nonumber\\
&&+z^iM^I(T_I)_{ij^*}\eta^{j^*k}M^J(T_J)_{kl^*}
\overline z^{l^*}\nonumber\\
&&+2\a g_{IJ}M^IP^J+\zeta^{\tilde I}{\cal C}_{\tilde I}^{\ I}g_{IJ}P^J
\end{eqnarray}
From the variation of the lagrangian with respect to the auxiliary
fields $H^i$ and $P^I$ we find:
\begin{eqnarray}
H^i&=&\eta^{ij^*}\partial_{j^*}\overline W(\overline z)\,,\\
P^I&=&D^I(z,\overline z)+2\a M^I+\zeta^{\tilde I}{\cal C}_{\tilde I}^{\ I}\,,
\end{eqnarray}
where
\begin{equation}
D^I(z,\overline z)=-\overline z^{i^*}(T_I)_{i^*j}z^j
\end{equation}
Substituting this expressions in the potential (\ref{defU}) we obtain:
\begin{eqnarray}
U(z,\overline z,M)&=&
-\partial_i W(z)\eta^{ij^*}\partial_{j^*}\overline W(\overline z)\nonumber\\
&&+\ft{1}{2}g^{IJ}\left(\overline z^{i^*}(T_I)_{i^*j}z^j\right)
\left(\overline z^{k^*}(T_J)_{k^*l}z^l\right)\nonumber\\
&&+\overline z^{i^*}M^I(T_I)_{i^*j}\eta^{jk^*}M^J(T_J)_{k^*l}
z^l\nonumber\\
&&-2\a^2g_{IJ}M^IM^J-2\a\zeta^{\tilde I}{\cal C}_{\tilde I}^{\ I}g_{IJ}M^J
-\ft{1}{2}\zeta^{\tilde I}{\cal C}_{\tilde I}^{\ I}g_{IJ}
\zeta^{\tilde J}{\cal C}_{\tilde J}^{\ J}\nonumber\\
&&-2\a M^I\left(\overline z^{i^*}(T_I)_{i^*j}z^j\right)
-\zeta^{\tilde I}{\cal C}_{\tilde I}^{\ I}
\left(\overline z^{i^*}(T_I)_{i^*j}z^j\right)
\end{eqnarray}
%
%
\subsection{A particular ${\cal N}=2$ theory: ${\cal N}=4$}
A general lagrangian for matter coupled rigid ${\cal N}=4, d=3 $
super Yang Mills theory is easily
obtained from the dimensional reduction of the
${\cal N}=2,d=4$ gauge theory (see \cite{BertFre}).
The bosonic sector of this latter lagrangian   is the following:
\begin{eqnarray}\label{N=4Lag}
\LL_{bosonic}^{{\cal N}=4}&=&-\frac{1}{g^2_{_{YM}}}g_{IJ}F^I_{mn}F^{J\,mn}
+\frac{1}{2g^2_{_{YM}}}g_{IJ}\nabla_m M^I\nabla^m M^J\nonumber\\
&&+\frac{2}{g^2_{_{YM}}}g_{IJ}\nabla_m\overline Y^I\nabla^mY^J
+\frac{1}{2}Tr\left(\nabla_m\overline{\bf Q}
\nabla^m{\bf Q}\right)\nonumber\\
&&-\frac{1}{g^2_{_{YM}}}g_{IN}f^I_{JK}f^N_{LM}\,M^J\overline Y^K\,M^L Y^M
-M^IM^JTr\left(\overline{\bf Q}(\hat T_I\,\hat T_J)
{\bf Q}\right)\nonumber\\
&&-\frac{2}{g^2_{_{YM}}}g_{IN}f^I_{JK}f^N_{LM}\,\overline Y^JY^K\,
\overline Y^LY^M
-\overline Y^IY^J\,Tr\left(\overline{\bf Q}\left\{\hat T_I,\hat T_J\right\}
{\bf Q}\right)
\nonumber\\
&&-\frac{1}{4}g^2_{_{YM}}g_{IJ}Tr\left(\overline{\bf Q}(\hat T^I){\bf Q}\,
\overline{\bf Q}(\hat T^J){\bf Q}\right)
\end{eqnarray}
The bosonic matter field content is given by two kinds of fields.
First we have a complex field $Y^I$ in the adjoint representation
of the gauge group, which belongs to a chiral multiplet.
Secondly, we have an $n$-uplet of quaternions
${\bf Q}$, which parametrize a (flat)\footnote{ Once again we choose
the  HyperK\"ahler manifold to be flat since we are interested in microscopic
theories with canonical kinetic terms} HyperK\"ahler manifold:
\begin{equation}
{\bf Q}=\left(\begin{array}{ccl}
Q^1&=&q^{1|0}\unity-iq^{1|x}\s_x\\
Q^2&=&q^{2|0}\unity-iq^{2|x}\s_x\\
&\cdots&\\
Q^A&=&q^{A|0}\unity-iq^{A|x}\s_x\\
&\cdots&\\
Q^n&=&q^{n|0}\unity-iq^{n|x}\s_x
\end{array}\right)
\qquad\begin{array}{l}
q^{A|0},q^{A|x}\in\IR\\
\\
A\in\{1,\ldots,n\}\\
\\
x\in\{1,2,3\}
\end{array}
\end{equation}
The quaternionic conjugation is defined by:
\begin{equation}
\overline Q^A=q^{A|0}\unity+iq^{A|x}\s_x
\end{equation}
In this realization, the quaternions are represented by
matrices of the form:
\begin{equation}
Q^A=\left(\begin{array}{cc}
u^A&i\overline v_{A^*}\\
iv_A&\overline u^{A^*}
\end{array}\right)\qquad
\overline Q^A=\left(\begin{array}{cc}
\overline u^{A^*}&-i\overline v_{A^*}\\
-iv_A&u^A
\end{array}\right)\qquad\begin{array}{l}
u^A=q^{A|0}-iq^{A|3}\\
V^m=-q^{A|1}-iq^{A|2}
\end{array}
\end{equation}
The generators of the gauge group ${\cal G}$ have a triholomorphic
action on the flat HyperK\"ahler manifold, namely they respect the
three complex structures. Explicitly this triholomorphic action
 on {\bf Q} is the following:
\begin{eqnarray}
\d^I{\bf Q}&=&i\hat T^I{\bf Q}\nonumber\\
&&\nonumber\\
\d^I\left(\begin{array}{cc}
u^A&i\overline v_{A^*}\\
iv_A&\overline u^{A^*}
\end{array}\right)&=&
i\left(\begin{array}{cc}
T^I_{A^*B}&\\
&-\overline T^I_{AB^*}
\end{array}\right)\left(\begin{array}{cc}
u^B&i\overline v_{B^*}\\
iv_B&\overline u^{B^*}
\end{array}\right)
\end{eqnarray}
where the $T^I_{A^*B}$ realize a representation
of ${\cal G}$ in terms of $n\times n$ hermitian matrices.
We define $\overline T_{AB^*}\equiv\left(T_{A^*B}\right)^*$, so,
being the generators hermitian ($T^*=T^T$), we can write:
\begin{equation}
T_{A^*B}=\overline T_{BA^*}.
\end{equation}
We can rewrite eq. (\ref{N=4Lag}) in the form:
\begin{eqnarray}\label{N=4bosonicLag}
\LL_{bosonic}^{{\cal N}=4}&=&-\frac{1}{g^2_{_{YM}}}g_{IJ}F^I_{mn}F^{J\,mn}
+\frac{1}{2g^2_{_{YM}}}g_{IJ}\nabla_m M^I\nabla^m M^J\nonumber\\
&&+\frac{2}{g^2_{_{YM}}}g_{IJ}\nabla_m\overline Y^I\nabla^m Y^J
+\nabla_m\overline u\nabla^m u+\nabla_m\overline v\nabla^mv\nonumber\\
&&-\frac{2}{g^2_{_{YM}}}M^I M^J\overline Y^R f_{RIL}f^L_{\,JS}Y^S
-M^I M^J\left(\overline u T_IT_J u
+\overline v\overline T_I\overline T_J v\right)\nonumber\\
&&-\frac{2}{g^2_{_{YM}}}g_{IJ}\left[\overline Y,Y\right]^I
\left[\overline Y,Y\right]^J
-2\overline Y^IY^J\left(\overline u\{T_I,T_J\} u
+\overline v\{\overline T_I,\overline T_J\} v\right)\nonumber\\
&&-2g^2_{_{YM}}g_{IJ}\left(vT^Iu\right)\left(\overline v\overline T^J
\overline u\right)
-\ft{1}{2}g^2_{_{YM}}g_{IJ}\left[\left(\overline u T^I u\right)
\left(\overline uT^Ju\right)\right.\nonumber\\
&&\left.+\left(\overline v\overline T^I v\right)\left(\overline v
\overline T^J v\right)-2\left(\overline u T^I u\right)\left(
\overline v\overline T^J v\right)\right]
\end{eqnarray}
By comparing the bosonic part of (\ref{N=2stLag})
(rescaled by a factor ${4\over g_{_{YM}}^2}$) with (\ref{N=4bosonicLag}),
we see that in order for a ${\cal N}\!\!=\!\!2$ lagrangian to be also
${\cal N}\!\!=\!\!4$ supersymmetric, the matter content of the theory
and the form of the superpotantial are constrained.
The chiral multiplets have to be in an adjoint plus a
generic quaternionic representation of $\cal G$.
So the fields $z^i$ and the gauge generators are
\begin{equation}
z^i=\left\{\begin{array}{l}
\sqrt 2Y^I\\
g_{_{YM}}u^A\\
g_{_{YM}}v_A
\end{array}\right.\qquad
T^I_{i^*j}=\left\{\begin{array}{l}
f^I_{\,JK}\\
(T^I)_{A^*B}\\
-(\overline T^I)_{AB^*}
\end{array}\right.\,.
\end{equation}
Moreover, the holomorphic superpotential $W(z)$ has to be of the form:
\begin{equation}
W\left(Y,u,v\right)=2g^4_{_{YM}}\delta^{AA^*}Y^I\,v_A(T_I)_{A^*B}u^B\,.
\end{equation}
Substituting this choices in the supersymmetric lagrangian
(\ref{N=2stLag}) we obtain the general ${\cal N}\!\!=\!\!4$ lagrangian
expressed in ${\cal N}\!\!=\!\!2$ language.
\par
Since the action of the gauge group is triholomorphic there is a
triholomorphic momentum map associated with each gauge group
generator (see \cite{ALE,damia,BertFre})
\par
The momentum map is given by:
\begin{equation}
{\cal P}=\ft{1}{2}i\left(\overline{\bf Q}\,\hat T\,{\bf Q}\right)=
\left(\begin{array}{cc}
{\cal P}_3&{\cal P}_+\\
{\cal P}_-&-{\cal P}_3
\end{array}\right)\,,
\end{equation}
where
\begin{eqnarray}
{\cal P}_3^I&=&-\left(\overline uT^Iu-\overline v\overline T^Iv\right)=
D^I\nonumber\\
{\cal P}_+^I&=&-2\overline v\overline T^I\overline u=
-g^{-4}_{_{YM}}\ \partial\overline W/\partial \overline Y_I\nonumber\\
{\cal P}_-^I&=&2vT^Iu=
g^{-4}_{_{YM}}\ \partial W/\partial Y_I\,.
\end{eqnarray}
So the superpotential can be written as:
\begin{equation}
W=g^4_{_{YM}}Y_I{\cal P}_-^I\,.
\end{equation}
%
%
\subsection{A particular ${\cal N}=4$ theory: ${\cal N}=8$}
In this section we discuss the further conditions under which the
${\cal N}\!\!=\!\!4$ three dimensional lagrangian previously
derived acquires an ${\cal N}\!\!=\!\!8$ supersymmetry.
To do that we will compare the four dimensional ${\cal N}\!\!=\!\!2$
lagrangian of \cite{BertFre} with the four dimensional ${\cal N}\!\!=\!\!4$
lagrangian of \cite{N8DF} (rescaled by a factor ${4\over g_{_{YM}}^2}$),
whose bosonic part is:
\begin{eqnarray}\label{N=8piphiLag}
{\cal L}^{{\cal N}=4~D=4}_{bosonic}&=&{1\over g_{_{YM}}^2}
\left\{-F^{\underline m\underline n}F_{\underline m\underline n}+
{1\over 4}\nabla^{\underline m}\phi^{AB}\nabla_{\underline m}\phi^{AB}+
{1\over 4}\nabla^{\underline m}\pi^{AB}\nabla_{\underline m}\pi^{AB}
\right.\nonumber\\
&+&{1\over 64}\left(
\left[\phi^{AB},\phi^{CD}\right]\left[\phi^{AB},\phi^{CD}\right]+
\left[\pi^{AB},\pi^{CD}\right]\left[\pi^{AB},\pi^{CD}\right]
\right.\nonumber\\
&+&\left.\left.2\left[\phi^{AB},\pi^{CD}\right]
\left[\phi^{AB},\pi^{CD}\right]\right)\right\}
\end{eqnarray}
The fields $\pi^{AB}$ and $\phi^{AB}$ are Lie-algebra valued:
\begin{equation}
\left\{\begin{array}{ccl}
\pi^{AB}&=&\pi^{AB}_It^I\\
\phi^{AB}&=&\phi^{AB}_It^I
\end{array}\right.\,,
\end{equation}
where $t^I$ are the generators of the gauge group $\cal G$.
They are the real and imaginary parts
of the complex field $\rho$:
\begin{equation}
\left\{\begin{array}{ccl}
\rho^{AB}&=&\ft{1}{\sqrt{2}}\left(\pi^{AB}+i\phi^{AB}\right)\\
\overline{\rho}_{AB}&=&\ft{1}{\sqrt{2}}\left(\pi^{AB}
-i\phi^{AB}\right)
\end{array}\right.\,.
\end{equation}
$\rho^{AB}$ transforms in the represention $\bf 6$ of a
global $SU(4)$-symmetry of the theory.
Moreover, it satisfies the following pseudo-reality condition:
\begin{equation}
\rho^{AB}=-\ft{1}{2}\e^{ABCD}\overline{\rho}_{CD}
\end{equation}
In terms of $\rho$ the lagrangian (\ref{N=8piphiLag}) can be
rewritten as:
\begin{equation}\label{N=8rhoLag}
{\cal L}^{{\cal N}=8}_{bosonic}={1\over g_{_{YM}}^2}\left\{
-F^{\underline m\underline n}F_{\underline m\underline n}
+{1\over 2}\nabla_{\underline m}\overline{\rho}_{AB}\nabla^{\underline m}
\rho^{AB}
+{1\over 16}\left[\overline{\rho}_{AB}\rho^{CD}\right]\left[\rho^{AB},
\overline{\rho}_{CD}\right]\right\}
\end{equation}
The $SU(2)$ global symmetry of the ${\cal N}\!\!=\!\!2,~D\!\!=\!\!4$
theory can be diagonally embedded into the $SU(4)$ of
the ${\cal N}\!\!=\!\!4,~D\!\!=\!\!4$ theory:
\begin{equation}
{\cal U}=\left(\begin{array}{cc}
U&0\\
0&\overline U
\end{array}\right)\ \in SU(2)\subset SU(4)\,.
\end{equation}
By means of this embedding, the $\bf 6$ of SU(4) decomposes as
${\bf 6}\longrightarrow{\bf 4+1+1}$.
Correspondingly, the pseudo-real field $\rho$ can be splitted into:
\begin{eqnarray}
\rho^{AB}&=&\left(\begin{array}{cccc}
0&\sqrt 2Y&g_{_{YM}} u&ig_{_{YM}}\overline v\\
-\sqrt 2Y&0&ig_{_{YM}} v&g_{_{YM}}\overline u\\
-g_{_{YM}} u&-ig_{_{YM}} v&0&-\sqrt 2\overline Y\\
-ig_{_{YM}}\overline v&-g_{_{YM}}\overline u&\sqrt 2\overline Y&0
\end{array}\right)\nonumber\\
&&\nonumber\\
&=&\left(\begin{array}{cc}
i\sqrt 2\s^2\otimes Y&g_{_{YM}}Q\\
&\\
-g_{_{YM}}Q^T&-i\sqrt 2\s^2\otimes\overline Y
\end{array}\right)\,,
\end{eqnarray}
where $Y$ and $Q$ are Lie-algebra valued.
The global $SU(2)$ transformations act as:
\begin{equation}
\rho\longrightarrow{\cal U}\rho{\cal U}^T=\left(\begin{array}{cc}
i\sqrt 2\s^2\otimes Y&g_{_{YM}}UQU^{\dagger}\\
&\\
-g_{_{YM}}\left(UQU^{\dagger}\right)^T&-i\sqrt 2\s^2\otimes\overline Y
\end{array}\right)
\end{equation}
Substituting this expression for $\rho$ into (\ref{N=8rhoLag})
and dimensionally reducing to three dimensions, we obtain the
lagrangian (\ref{N=4Lag}).
In other words the ${\cal N}\!\!=\!\!4,~D\!\!=\!\!3$ theory is
enhanced to ${\cal N}\!\!=\!\!8$ provided the hypermultiplets
are in the adjoint representation of $\cal G$.
\section{Conclusions}
\label{conclu}
In this paper we have discussed an essential intermediate step for
the comparison between Kaluza Klein supergravity
compactified on manifolds $AdS_4 \times X^7$ and
superconformal field theories living on
an M2 brane world volume. Focusing on the case with ${\cal N}=2$
supersymmetry we have shown how to convert Kaluza Klein data on
$Osp(2\vert 4)$ supermultiplets into conformal superfields living in
three dimensional  superspace. In addition since such conformal
superfields are supposed to describe composite operators of a
suitable $d=3$ gauge theory  we have studied the general form of
three dimesnional $N=2$ gauge theories. Hence in this paper we have
set the stage for the discussion of specific gauge theory models
capable of describing, at an infrared conformal point the Kaluza Klein
spectra, associated with the sasakian seven--manifolds (\ref{sasaki})
classified in the eighties and now under active consideration once again.
Indeed the possibility of constructing dual pairs \{$M2$--brane gauge
theory,supergravity on $G/H$ \} provides a challenging testing ground for the
exciting AdS/CFT correspondence.
\section{Acknowledgements}
We are mostly grateful to Sergio Ferrara for essential and
enlightening discussions and for introducing us to the superfield
method in the construction of the superconformal multiplets. We would
also like to express our gratitude to C. Reina, A. Tomasiello and A.
Zampa for very important and clarifying discussions on the
geometrical aspects of the AdS/CFT correspondence and to A. Zaffaroni
for many clarifications on brane dynamics. Hopefully
these discussions will lead us to the construction of the gauge
theories associated with the four sasakian manifolds of
eq.(\ref{sasaki}) in a joint venture.
\appendix
%
%
\vskip 2cm
\begin{center}
  {\Large {\bf Appendix}}
\end{center}
\section{Calculation of the Killing vectors}
\label{derivationkillings}
To evaluate
the left-hand-side of (\ref{defcoset}) which
has the form $e^A \, e^B$, we use the
Campell-Baker-Hausdorff formula for
$A$ an infinitesimal generator:
\begin{eqnarray}
e^A \, e^B =
\exp\Big(A + B +  z_1 [B, A ] + z_2 [B, [B, A]]
+  \cdots + z_n [B, [ \cdots [B, A] \cdots ]]
\Big)\,,
\label{CBH}
\nonumber \\
\end{eqnarray}
where
$z_0=1, z_1=-\ft12, z_2=\ft{1}{12}, z_3=0,  z_4= - \ft{1}{720} \cdots$
\footnote{The coefficients $z_n$ are determined recursively
by
\begin{equation}
\frac{1}{(n+1)!} + \frac{z_1}{n!}
+ \frac{z_2}{(n-1)!}+ \cdots + z_n = 0 \,.
\end{equation}}
We plug the equation (\ref{usefulcommutator}) iteratively into (\ref{CBH}).
Then one sees that for the operators $D$, $P_m$ and $q^{\alpha i}$,
 the compensators $h_{D}, h_P$ and $h_q$ are zero. We can determine
the complete expressions for their Killing vectors,
\footnote{We write $\left(\gamma\theta^i\right)^\alpha=
(\gamma^m)^\alpha{}_\beta\theta^{\beta i} $.}
\begin{eqnarray}
\veck [P_m] &=& -i\, \pi(\rho) \partial_m \,, \nonumber \\
\veck [D] &=& \frac{\partial}{\partial \rho} + \delta(\rho) \, x \cdot \partial
+ d(\rho) \, \theta^i_\alpha \frac{\partial}{\partial \theta^i_\alpha} \,,
\nonumber \\
\veck [q^{\alpha i}] &=& t(\rho) \frac{\partial}{\partial\theta^i_\alpha}
+  \g(\rho) \left(\gamma\theta^i\right)^\alpha \partial_m \,,
\end{eqnarray}
where,
\begin{eqnarray}
\pi(\rho) &=& \sum_{n=0}^\infty (-)^n z_n \rho^n \,,
\nonumber \\
\delta(\rho)&=&\sum_{n=1}^\infty (-)^{n-1} z_n \rho^{n-1} \,,
\nonumber \\
d(\rho) &=& \sum_{n=1}^\infty (-)^{n-1} \frac{1}{2^n} z_n \rho^{n-1}
\,, \nonumber \\
t(\rho) &=& \sum_{n=0}^\infty (-)^{n} \frac{1}{2^n} z_n \rho^{n}
\,, \nonumber \\
\g(\rho) &=& \sum_{n=1}^\infty (-)^{n-1} c_n z_n \rho^{n-1} \,,
\label{killingfunctions}
\end{eqnarray}
and
\begin{equation}
c_{n+1} = c_n +\frac{1}{2^n}\,, \qquad c_1=1 \,.
\end{equation}
These functions satisfy some differential relations that
are needed for the closure of the algebra. For example,
\begin{eqnarray}
\pi^\prime - \pi \delta &=& \pi \,,
\nonumber \\
-2\, t \g &=& \pi \,,
\nonumber \\
t^\prime - t  d &=& \ft12\, t \,,
\nonumber \\
\g^\prime + \g d - \g d &=& \ft12 \, \g \,,
\label{propertiesfunctions}
\end{eqnarray}
which ensures closure of the commutators $[D, P_m]$, $[D, q^{\alpha i}]$
and $\{q^{\alpha i}, q^{\beta j}\}$.
Upon taking an $SO(1,2)$-rotation for $g=e^{j \cdot J}$ in (\ref{defcoset})
one sees that the compensator does not vanish but equals $g$,
\begin{equation}
h_J(y) = e^{j \cdot J} \,.
\end{equation}
Consequently one finds the complete expression for the Killing vector of
the generator $J^m$,
\begin{eqnarray}
\veck [J^m] &=& \varepsilon^{mpq} x_p \partial_q
-\frac{i}{2} \left(\theta^i\gamma^m \right)_\alpha \frac{\partial}{\partial
\theta_\alpha^i}
\,.
\end{eqnarray}
Similarly, for the $SO({\cal N})$ group element $g=e^{t_{ij}\, T^{ij}}$ we find,
\begin{equation}
h_T = e^{t_{ij}\, T^{ij}}
\end{equation}
and
\begin{eqnarray}
\veck [T^{ij}]
&=& i \left( \theta_\alpha^i \frac{\partial}{\partial \theta_\alpha^j}
- \theta_\alpha^j \frac{\partial}{\partial \theta_\alpha^i} \right)
\,.
\end{eqnarray}
Notice that both for the
$SO(1,2)$ and the $SO({\cal N})$
the Killing vectors do not depend on the coordinate $\rho$.
Finding the superspace operators for $s_\alpha^i, K_m$
is more involved. We restrict ourselves here to find
the superspace operators for small $\rho$, i.e.
close to the boundary of $AdS$.
However we have checked that the expressions
for the Killing vectors (\ref{simplekilling}) and the
compensators (\ref{compensatorsAdSrzxi}) are complete.
The $s$-supersymmetry is needed to find extra constraints
on constrained superfields. There is no
need to consider the $K$-transformations.
A multiplet that transforms properly under
$s$-supersymmetry will also transform properly under
$K$-transformations since
\begin{equation}
K_m = \ft{i}{2} \, s^i \gamma_m s^i \,.
\label{Kiss2}
\end{equation}
The Killing vectors for small $\rho$ (first-order approximation) are
\begin{eqnarray}
\veck [P_m] &=& -i\, \left(1+\ft12 \rho\right)\partial_m  \,,
\nonumber \\
\veck [q^{\alpha i}] &=& \left(1+\ft14\rho\right)
\frac{\partial}{\partial \theta_\alpha^i}
-\frac{1}{2}\left(1+\ft14\rho\right) \left(\gamma^m\theta^i\right)^\alpha
\partial_m\,,
\nonumber \\
\veck [J^m] &=& \varepsilon^{mpq} x_p \partial_q
-\frac{i}{2} \left(\theta^i\gamma^m \right)_\alpha \frac{\partial}{\partial
\theta_\alpha^i}
\,,
\nonumber \\
\veck [D] &=& \frac{\partial}{\partial \rho}
-\frac{1}{2}\left(1+\ft16\rho\right) x \cdot \partial
-\frac{1}{4}\left(1+\ft{1}{12}\rho\right)\theta_\alpha^i
\frac{\partial}{\partial \theta_\alpha^i}
\,,
\nonumber \\
\veck [s^{\alpha i}] &=& -\left(1-\ft14 \rho\right) \theta^{\alpha i}
\frac{\partial}{\partial \rho} +
\frac{1}{12} \rho \, \theta^{\alpha i} x \cdot
\partial
 + \frac{i}{2} \left(1-\ft14\rho \right)
\varepsilon^{pqm} \, x_p \left(\gamma_q\theta^i\right)^\alpha \partial_m
\nonumber \\
& &
- \frac{1}{8} \left( 1 - \ft14 \rho \right) \, \theta^j \theta^j
\,  \left( \gamma^m \theta^i \right)^\alpha \, \partial_m
 - \left( 1 - \ft14 \rho \right) \,
x^m \, \left( \gamma_m \right)^\alpha{}_\beta
\frac{\partial}{\partial \theta^i_\beta}
\nonumber \\
& & +\frac{1}{4} \, \left( 1 - \ft16 \rho \right) \, \theta^{\alpha i}
\, \theta_\beta^j \frac{\partial}{\partial \theta^j_\beta}
 - \frac{1}{4} \left( 1 - \ft14 \rho \right) \, \theta^j \theta^j
 \frac{\partial}{\partial \theta^i_\alpha}
\nonumber \\
& & - \frac{1}{2} \left( 1 - \ft14 \rho \right)
\left( \gamma_m \theta^i \right)^\alpha
\, \theta^j \gamma^m \frac{\partial}{\partial \theta^j} \,,
\nonumber \\
\veck [K_m] &=& - \ft{i}{2} \, \veck [s^i] \, \gamma_m \veck [s^i]
\,,
\nonumber \\
\veck [T^{ij}] &=& i \left( \theta_\alpha^i \frac{\partial}{\partial \theta_\alpha^j}
- \theta_\alpha^j \frac{\partial}{\partial \theta_\alpha^i} \right)
\,.
\label{smallrhooperators}
\end{eqnarray}
and the compensators ($\rho<<1$) are given by
\begin{eqnarray}
W[P]&=& 0 \,, \nonumber \\
W[q^{\alpha i}] &=& 0 \,, \nonumber \\
W[J^m] &=& J^m \,, \nonumber \\
W[D] &=& 0 \,, \nonumber \\
W[s^{\alpha i}] &=& \left( 1-\ft12\rho \right) \, s^{\alpha i}
  - i\, \left( 1-\ft14\rho \right) \,
\left(\gamma^m \theta^i \right)^\alpha \,J_m
  + i \left(1-\ft14\rho\right) \theta^{\alpha j} \, T^{ij} \,.
\nonumber \\
\label{compensatorsAdS}
\end{eqnarray}
Using the functions (\ref{killingfunctions})
it turns out that the change of coordinates
from the $(\rho,x,\theta)$-basis to the $(\rho,z,\xi)$-basis
can actually be written as
\begin{eqnarray}
x &=& \pi(\rho) \, z \,, \nonumber \\
\theta &=& t(\rho) \, \xi \,.
\end{eqnarray}
Using the properties (\ref{propertiesfunctions}) one sees that indeed
the bounary Killing vectors get the simple form of (\ref{simplekilling}).
\par
Treating the boundary as a different coset, one understands that
the superspace operators a priori are not retrieved by just putting
$\rho=0$ in the operators (\ref{smallrhooperators}).
To illustrate this, let us look at the dilatation.
Let us take $g=e^{\rho\,D}$, then
 on the coset $G/{H^{CFT}}$ we need a compensator $h_D^{CFT}=g$
and find
\begin{equation}
kD\Big\vert_{\partial AdS} =
- x \cdot \partial
-\frac{1}{2}\theta_\alpha^i
\frac{\partial}{\partial \theta_\alpha^i} \,,
\end{equation}
and one sees that this is not the Killing vector $kD$
in the parametrization
$(\rho, x ,\theta)$ that acts on the coset representative
(\ref{AdScosetrepresentative}) with $\rho=0$.
Yet this is clearly the case for the $(\rho, z , \theta)$
parametrization. And hence the parametrization
(\ref{simplecostrepresentative}) is the most
suitable for a comparative study between the boundary
and the bulk of $AdS$ superspace.
%
%
\section{Useful identities}
\begin{eqnarray}
\gamma^n \gamma^m \gamma_n = - \gamma^m
\end{eqnarray}
Fierz relation,
\begin{eqnarray}
\phi^\alpha \psi_\beta = -\ft12 \, \delta^\alpha{}_\beta \, \psi \phi
-  \ft12 \, (\gamma^m)^\alpha{}_\beta  \, \psi \gamma_m \phi \,,
\nonumber \\
\left(\gamma^m\right)^\alpha{}_\beta \left(\gamma_m\right)^\gamma{}_\delta =
- \delta^\alpha{}_\beta \delta^\gamma{}_\delta +  2 \, \delta^\alpha{}_\delta
\delta^\gamma{}_\beta \,.
\end{eqnarray}
The following identity can be used to plug it iteratively in
formula (\ref{CBH}),
\begin{eqnarray}
&&\left[\rho \,D + i\, x \cdot P + \theta^i q^i \,,
R\, D + X \cdot P + \Theta^j  q^j + L \cdot K
+ j \cdot J + \Sigma^j  s^j + t_{jk} \, T^{jk}
{}\right]
\nonumber \\
&& \hskip 3 cm = \nonumber\\
&&\left(-2i  \, x \cdot L + \Sigma^i \theta^i  \right) \, D
\nonumber \\
&&
+\left( - \rho \, X^m + i\, R \, x^m + i\, \Theta^i \gamma^m \theta^i
+i\, \varepsilon^{pqm}\, x_p \, j_q \right) \, P_m
\nonumber \\
&&
+ \left( -\ft12 \, \rho \,\Theta_\alpha^i + \ft12 \, R \, \theta_\alpha^i
+ \ft{i}{2} \, j^m \left(\theta \gamma_m\right)_\alpha
- 2\, i \, t_{ji} \, \theta_\alpha^j
+ x^m \left(\Sigma^i  \gamma_m \right)_\alpha \right) q^{\alpha i}
\nonumber \\
&&
+ \left(\rho \, L^m \right) K_m
\nonumber \\
&&
+ \left( 2i \, \varepsilon^{pqm}\, L_p \, x_q
+ i\, \Sigma^i \gamma^m \theta^i \right) J_m
\nonumber \\
&&
+ \left( \ft12 \, \rho \, \Sigma_\alpha^i
- i\, L^m \left(\theta^i \gamma_m\right)_\alpha \right) s^{\alpha i}
\nonumber \\
&&
+\left( -i\, \Sigma^{[i} \theta^{j]} \right) T^{ij}
\,.
\label{usefulcommutator}
\end{eqnarray}
%
%
\end{document}